\documentclass {IEEEtran}
\usepackage{bm}
\usepackage{epsfig}
\usepackage{stfloats}
\usepackage{graphicx}
\usepackage{subfigure}
\usepackage{epstopdf}
\usepackage{multirow}
\usepackage[sort, compress]{cite}
\usepackage{epsfig,graphics,subfigure}
\usepackage[cmex10]{amsmath}

\usepackage{array}
\usepackage{tabulary}
\usepackage{booktabs}
\usepackage{amsfonts}
\usepackage{algorithmic}
\usepackage{algorithm}

\usepackage{epstopdf}

\usepackage{amssymb}

\usepackage{textcomp}
\usepackage{xcolor}

\def\BibTeX{{\rm B\kern-.05em{\sc i\kern-.025em b}\kern-.08em
    T\kern-.1667em\lower.7ex\hbox{E}\kern-.125emX}}

\begin{document}
\bibliographystyle{IEEEtran}

\title{Movable Frequency Diverse Array-Assisted Covert Communication With Multiple Wardens}
\author{\IEEEauthorblockN{Zihao Cheng, Jiangbo Si, {\emph{Member, IEEE}}, Zan Li, {\emph{Senior Member, IEEE}}, \\
Pengpeng Liu, Xiaoting Wang and Naofal Al-Dhahir, \emph{Fellow, IEEE}}}
 \maketitle
\begin{abstract}
The frequency diverse array (FDA) is highly promising for improving covert communication performance by adjusting the frequency of each antenna at the transmitter. However, when faced with the cases of multiple wardens and highly correlated channels, FDA is limited by the frequency constraint and cannot provide satisfactory covert performance. In this paper, we propose a novel movable FDA (MFDA) antenna technology where positions of the antennas can be dynamically adjusted in a given finite region. Specifically, we aim to maximize the covert rate by jointly optimizing the antenna beamforming vector, antenna frequency vector and antenna position vector. To solve this non-convex optimization problem with coupled variables, we develop a two-stage alternating optimization (AO) algorithm based on the block successive upper-bound minimization (BSUM) method. Moreover, considering the challenge of obtaining perfect channel state information (CSI) at multiple wardens, we study the case of imperfect CSI. Simulation results demonstrate that MFDA can significantly enhance covert performance compared to the conventional FDA. In particular, when the frequency constraint is strict, MFDA can further increase the covert rate by adjusting the positions of antennas instead of the frequencies.

\end{abstract}
\begin{IEEEkeywords}
Covert communication, Frequency diverse array, Movable antenna, Phased array.
\end{IEEEkeywords}
\IEEEpeerreviewmaketitle

\section{Introduction}
Covert communication is a promising technology that has attracted extensive attention in recent years. Many techniques have been proposed to enhance the covert performance, where multi-antenna technique exploit the spatial degrees of freedom (DoFs) in transmit antennas to focus the power of signals toward the legitimate receiver and away from the warden by smartly designing the beamformer \cite{8654724,9438645,8959086,9099441,10090449}. In particular, based on the global perfect Channel State Information (CSI), the signals can achieve spatial cancellation without being detected by the warden \cite{10090449}. It is worth mentioning that this beamforming scheme for covert communication relies on phased array (PA) antennas. Hence, the improvement of covert performance requires that the legitimate receiver and the adversary is not locate in the same beam direction. However, in realistic scenarios, the warden may approach the legitimate receiver and be in the same beam direction, which leads to a degradation in covert performance.

Frequency diverse array (FDA) is a promising antenna technology used at the multi-antenna transmitter in wireless communization systems \cite{7087373,7817778,7084678,6951408,1631800}. Compared with conventional PA, FDA introduces a small frequency increment across the array elements to produce distinct accumulating phase lags among array antennas and forms a special beampattern related to distance and angle. By tuning the antenna frequency vector (AFV) at the transmitter, FDA can enable the transmitted signals to constructively superimpose at the desired positions and destructively cancel at the undesired positions simultaneously \cite{7084678,6951408,1631800}. Especially in covert communication, FDA can achieve perfect covertness by optimizing AFV with the instantaneous channel state information (CSI) at the warden \cite{10097703,9905724}. Moreover, FDA develops an angle-range dependent beampattern which helps decouple the correlated channels to enhance covert performance. In particular, by optimizing AFV, the received signal levels of the legitimate user and the warden can be distinguished clearly even if they are in the same beam direction. In the multi-warden scenario, a larger frequency increment is needed to decouple the correlated channels \cite{9133276,9560277}. However, the frequency increment is usually assumed to be much smaller than the carrier frequency due to receiver restrictions which greatly limits FDA performance.

It is worth mentioning that both FDA and PA are fixed-position arrays which means that the distance between each antenna is fixed and no smaller than a half-wavelength \cite{6798744,6736761,7416205}. Recently, movable antenna (MA) technology was proposed to achieve better communication performance where the antennas at the transmitter are movable within a confined region \cite{10243545,10286328,10318061,10278220,10304448,10354003,10464791,10382559}. By smartly designing the antenna positions vector (APV), MA can achieve higher spatial diversity gains by employing much fewer antennas compared with the conventional antenna selection (AS) technology \cite{10243545,10286328,10318061}. Specifically, with known global instantaneous CSI, MA can achieve full array gain with null steering \cite{10278220}. Moreover, MA is capable of mitigating the interference from an adversary when it is deployed at the receiver \cite{9369091}. In a multi-user and multi-interference scenario, MA can move in a three-dimensional (3D) region to obtain higher rate gain from the users and reduce the impact from different interference sources simultaneously. In addition, MA has great flexibility which can be deployed in one-dimensional (1D) line, two-dimensional (2D) surface and even six-dimensional (6D) region \cite{10304448,10354003,10464791,10382559}.

In light of the above considerations, we propose a novel antenna technology, namely movable frequency diverse array (MFDA), which can further exploit the spatial DoFs in transmit antennas based on FDA. Our main contributions are summarized as follows:


\begin{itemize}
\item[1.]

MFDA-aided covert communication with multiple wardens is firstly investigated in this paper. Different from FDA, MFDA can not only adjust the frequency of each antenna but also adjust the position, such that it achieves two-dimensional security. Specifically, we jointly optimize the antenna beamforming vector (ABV), AFV and APV at the transmitter to maximize the covert rate subject to the given constraints. Moreover, we investigate the effects of imperfect CSI at wardens on covert performance and the case of perfect covertness.

\item[2.]

Since the formulated optimization problem is non-convex with multiple coupled variables, we propose a low-complexity two-stage alternating optimization (AO) algorithm to solve this problem. Specifically, in the first stage, we alternately optimize AFV and APV by utilizing the block successive upper-bound minimization (BSUM) method. In the second stage, with optimized AFV and APV, the optimization problem is convex and can be solved by standard convex optimization solvers, e.g., CVX. For the case of imperfect CSI, we assume that the exact coordinates of wardens are unknown, but the uncertain regions of wardens are known. We first divide the regions into multiple discrete samples and consider the sum of the channel correlations. Then, the proposed two-stage AO algorithm can be utilized to solve the problem.

\item[3.]

MFDA can achieve better covert performance than FDA and PA. In particular, when the channels of the legitimate user and wardens are highly correlated, FDA is limited by the frequency constraint and cannot provide satisfactory performance. By contrast, MFDA can adjust the positions of antennas instead of the frequencies to further improve covert performance. Moreover, for the case of imperfect CSI, MFDA can achieve signal covertness in the uncertain regions.

\end{itemize}

The rest of this paper is organized as follows. Section II describes the MFDA-aided covert system model and formulates the optimization problem.  Section III develops a two-stage AO algorithm to solve the problem with perfect CSI at wardens. Section VI considers the case of imperfect CSI at wardens. Numerical results are presented and discussed in Section V, and conclusions are drawn in Section IV.

\emph{Notations}: Boldface uppercase and lowercase letters denote matrices and vectors, respectively. Italic letters denote scalars; ${\bf{x}}^\dag $ denotes the Hermitian of a vector  $\bf{x} $. For a given matrix $\bf{A}$, we denote its transpose, Hermitian and inverse by ${\bf{A}}^T$, ${\bf{A}}^{\dag}$ and ${\bf{A}}^{-1}$, respectively. Denote ${\left\| . \right\|^2}$ as ${\ell _2}$-norm. $\mathbb{C}^{M \times N}$ denotes the complex $M \times N$ matrices. $ \left\langle { \cdot , \cdot } \right\rangle $ denotes the inner product. ${\bf{tr}} \left(  \cdot  \right)$ is trace operator of a matrix. $\left\lfloor  \cdot  \right\rfloor $ and $\left\lceil  \cdot  \right\rceil $ are the nearest integer towards $ - \infty $ and $ + \infty $, respectively. $c = a{\kern 1pt} \bmod {\kern 1pt} b$ means $c$ is the remainder of integer $a$ divided by integer $b$.

\section{System Model}

\begin{figure}[htbp]
\centering
\subfigure[]{
\includegraphics[width=1.5in]{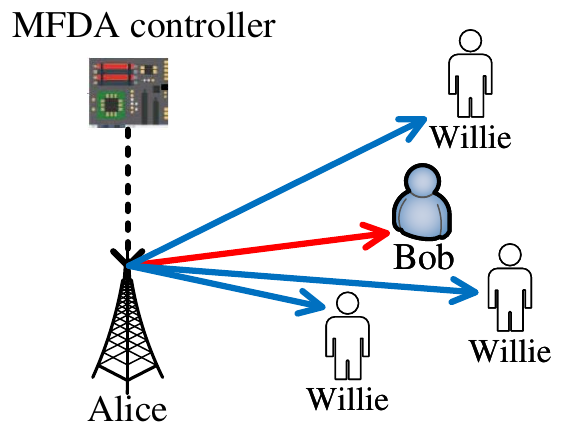}
\label{Fig1-1}}
\quad
\subfigure[]{
\includegraphics[width=1.5in]{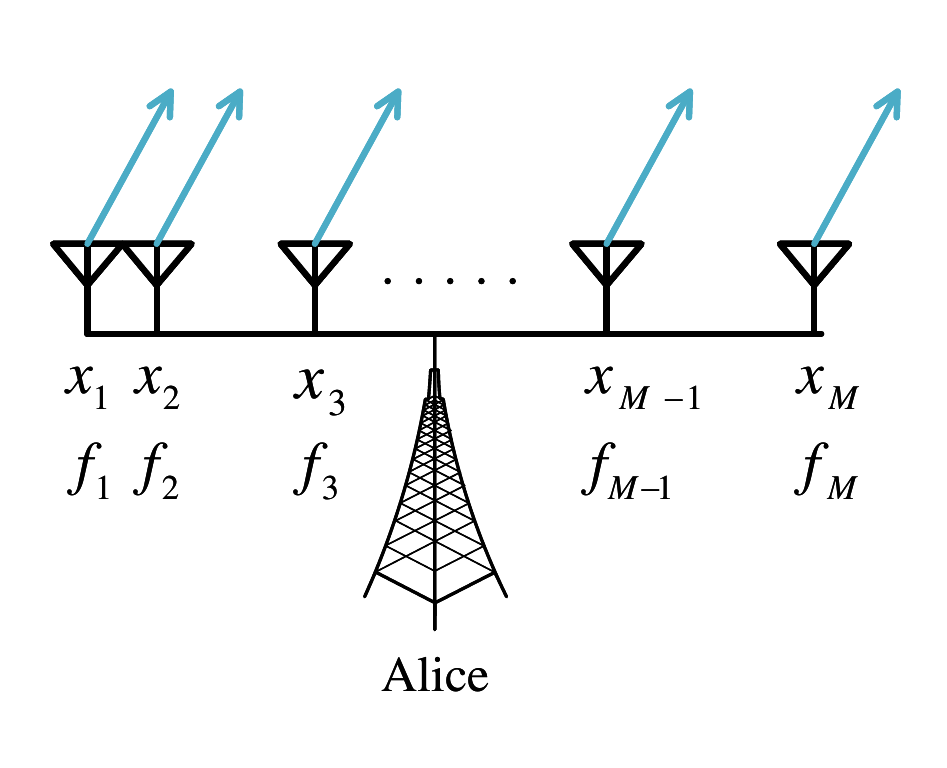}
\label{Fig1-2}}
\caption{MFDA-aided covert communication system model}\label{Fig1}
\end{figure}

\subsection{System Description and Channel Model}

In this paper, we consider a covert communication system with multiple wardens where a 1D linear MFDA is deployed to assist in the communication from the legitimate user Alice to Bob, while preventing being detected by $K$ wardens Willies. The number of antennas at Alice is denoted by $M$, and both Bob and Willies are assumed to be equipped with a single antenna. As shown in Fig. 1(a), a smart MFDA controller helps adjust the APV ${\bf{x}}\buildrel \Delta \over = {\left[ {{x_1},...,{x_M}} \right]} \in {{\mathbb{C}}^{1 \times M}}$ and the AFV ${\bf{f}} \buildrel \Delta \over = {\left[ {{f_1},...,{f_M}} \right]} \in {{\mathbb{C}}^{1 \times M}}$ at Alice via a separate wireless link. Compared with PA, MFDA introduces a small frequency increment across the array antennas, such that the frequency of the $m$-th antenna is given by
\begin{align}\label{radiation_frequency}\
{f_m} = {f_C} + \Delta {f_m},m = 1,...,M,
\end{align}
where $f_C$ is the carrier frequency and $\Delta {f_m}$ is the introduced frequency increment. We assume that $0 \le \Delta {f_m} \le \Delta F$ and the frequency increment is much smaller than the carrier frequency $\Delta F \ll {f_C}$.

Without loss of generality, we define the first antenna of Alice as the origin of the coordinate system. Thus, the coordinates of Bob and the $k$-th Willie are denoted by $\left( {{r_b},{\theta _b}} \right)$ and $\left( {{r_{{w_k}}},{\theta _{{w_k}}}} \right)$, respectively. To characterize the theoretical performance gain brought by the MFDA, we assume that $\left( {{r_b},{\theta _b}} \right)$ and $\left( {{r_{{w_k}}},{\theta _{{w_k}}}} \right)$ are perfectly known by Alice and all channels are line-of-sight (LoS) channels as in \cite{8078202,8104957,8458419}. In addition, the channels of Alice$\to$Bob and Alice$\to$ the $k$-th Willie are defined as
\begin{align}\label{h_ab}\
{{\bf{h}}_{ab}} = {\rm{Lfs}}\left( {{r_b}} \right)\left[ {{h_1}\left( {t,{r_b},{\theta _b}} \right),...,{h_M}\left( {t,{r_b},{\theta _b}} \right)} \right]
\end{align}
and
\begin{align}\label{h_ae}\
{{\bf{h}}_{a{w_k}}} = {\rm{Lfs}}\left( {{r_{{w_k}}}} \right)\left[ {{h_1}\left( {t,{r_{{w_k}}},{\theta _{{w_k}}}} \right),...,{h_M}\left( {t,{r_{{w_k}}},{\theta _{{w_k}}}} \right)} \right],
\end{align}
where ${h_m}\left( {t,r,\theta } \right) = {e^{ - j2\pi {f_m}\left[ {t - \frac{{r - {x_m}\sin \theta }}{c}} \right]}}$, $\forall m$. $c$ is the speed of light and ${\rm{Lfs}}\left( \cdot  \right)$ is the path loss factor.

\subsection{Covert Constraint}

We observe from \eqref{h_ab} and \eqref{h_ae} that the channels of MFDA are dependent on time $t$. Thus, the ABV at Alice is assumed to be time-varying to adapt to the variable channels which are defined as
\begin{align}\label{beamforming}\
{\bf{w}}\left( t \right) = {\left[ {{w_1}\left( t \right),...,{w_M}\left( t \right)} \right]^T}.
\end{align}
From Willies' perspective, a binary hypothesis testing problem should be considered under two events:
${\cal{H}}_0$ denotes the null hypothesis when Alice does not transmit information, and ${\cal{H}}_1$
denotes the alternative hypothesis when Alice transmits information. Thus, the received signal at the $k$-th Willie is given by
\begin{align}\label{received_willie}\
{y_{{w_k}}} = \left\{ {\begin{array}{*{20}{l}}
{{{\bf{n}}_{{w_k}}}}&{{{\cal{H}}_0}}\\
{{{\bf{h}}_{{aw_k}}}{\bf{w}}\left( t \right) + {{\bf{n}}_{{w_k}}}}&{{{\cal{H}}_1},}
\end{array}} \right.
\end{align}
where ${{\bf{n}}_{{w_k}}}$ is the additive white Gaussian noise (AWGN) at the $k$-th Willie with mean zero and variance $\sigma _{{w_k}}^2$, i.e., ${{\bf{n}}_{{w_k}}} \sim {\cal{CN}}\left( {0,\sigma _{{w_k}}^2}\right)$. To determine whether Alice transmits information or not, the $k$-th Willie needs to distinguish the probability curves of ${y_{w_k}}$ under ${\cal{H}}_0$ and ${\cal{H}}_1$. Based on \eqref{received_willie}, let ${p_{k,0}}\left( {{y_{w_k}}} \right)$ and ${p_{k,1}}\left( {{y_{w_k}}} \right)$ denote the likelihood functions of the received signals at the $k$-th Willie under ${\cal{H}}_0$ and ${\cal{H}}_1$, respectively. Now, ${p_{k,0}}\left( {{y_{w_k}}} \right)$ and
${p_{k,1}}\left( {{y_{w_k}}} \right)$ are given by
\begin{align}\label{H0}\
{p_{k,0}}\left( {{y_{{w_k}}}} \right) = \frac{1}{{\pi {\lambda _{k,0}}}}\exp \left( { - \frac{{{{\left| {{y_{{w_k}}}} \right|}^2}}}{{{\lambda _{k,0}}}}} \right)
\end{align}
and
\begin{align}\label{H1}\
{p_{k,1}}\left( {{y_{w_k}}} \right) = \frac{1}{{\pi {{\lambda _{k,1}} }}}\exp \left( { - \frac{{{{\left| {{y_{w_k}}} \right|}^2}}}{{{{\lambda _{k,1}} }}}} \right),
\end{align}
where ${\lambda _{k,0}} \buildrel \Delta \over = \sigma _{w_k}^2$ and ${\lambda _{k,1}} \buildrel \Delta \over = {\left| {{{\bf{h}}_{{aw_k}}}{\bf{w}}\left( t \right)} \right|^2} + \sigma _{{w_k}}^2$.

In general, the detection error probability (DEP) ${\xi}$ metric is used to evaluate covert performance. For an optimal detector at the $k$-th Willie, ${\xi_k}$ is
given by \cite{8714018,9381240}
\begin{align}\label{DEP}\
\xi_k  = 1 - {{\cal{V}}_T}\left( {{{p_{k,0}}},{{p_{k,1}}}} \right),
\end{align}
where ${{\cal{V}}_T}\left( {{{p_{k,0}}},{p_{k,1}}} \right)$ is the total variation between ${{p_{k,0}}}\left( {{y_{w_k}}} \right)$ and ${p_{k,1}}\left( {{y_{w_k}}} \right)$. Since computing
${{\cal{V}}_T}\left( {{{p_{k,0}}},{p_{k,1}}} \right)$ is intractable, we adopt Pinsker's inequality to obtain the following upper bound \cite{9398675}
\begin{align}\label{KL}\
{V_T}\left( {{{p_{k,0}}},{p_{k,1}}} \right) \le \sqrt {\frac{1}{2}{\cal{D}}\left( {{{p_{k,0}}}||{p_{k,1}}} \right)},
\end{align}
where ${{\cal{D}}\left( {{{p_{k,0}}}||{p_{k,1}}} \right)}$ is the Kullback-Leibler (KL) divergence, which is given by
\begin{align}\label{KL1}\
{\cal{D}}\left( {{p_{k,0}}||{p_{k,1}}} \right) &= \int_{ - \infty }^{ + \infty } {{p_{k,0}}\left( {{y_{{w_k}}}} \right)} \ln \frac{{{p_{k,0}}\left( {{y_{{w_k}}}} \right)}}{{{p_{k,1}}\left( {{y_{{w_k}}}} \right)}}dy\nonumber\\
 &= \ln \frac{{{\lambda _{k,1}}}}{{{\lambda _{k,0}}}} + \frac{{{\lambda _{k,0}}}}{{{\lambda _{k,1}}}} - 1.
\end{align}
Then, the covert constraint at the $k$-th Willie can be written as
\begin{align}\label{covert_constraint}\
{\cal{D}}\left( {{p_{k,0}}||{p_{k,1}}} \right) \le 2{\epsilon ^2},
\end{align}
where $ \epsilon$ is a small value and predetermined to ensure satisfactory covertness. According to \eqref{KL1}, $y\left( x \right) = \ln \left( x \right) + \frac{1}{x} - 1$ for $x \ge 1$ is a monotonically increasing function. Thus, the covert constraint of the $k$-th Willie can be simplified as
\begin{align}\label{covert_constraint_best}\
\frac{{{\lambda _{k,1}}}}{{{\lambda _{k,0}}}} \le {y^{ - 1}}\left( {2{\epsilon ^2}} \right),
\end{align}
where ${y^{ - 1}}\left( x \right)$ is the inverse function of $y\left( x \right)$.

\subsection{Problem Formulation}

The received signal at Bob can be written as
 \begin{align}\label{y_b}\
{{\bf{y}}_b}\left( t \right) = {{\bf{h}}_{ab}}{\bf{w}}\left( t \right) + {{\bf{n}}_b},
\end{align}
where ${{\bf{n}}_b}$ is the AWGN at Bob with mean zero and variance $\sigma _b^2$, i.e., ${{\bf{n}}_b} \sim {\cal{CN}}\left( {0,\sigma _b^2}\right)$. The covert rate at Bob is given by
 \begin{align}\label{R_b}\
{R_b}\left( t \right) = \log \left( {1 + \sigma _b^{ - 2}{{\left| {{\bf{h}}_{ab}{\bf{w}}\left( t \right)} \right|}^2}} \right).
\end{align}

To improve covert performance, we aim to maximize the covert rate by jointly optimizing ABV, APV and AFV at Alice, while avoiding being detected by different Willies. The optimization problem can be formulated as follows
\begin{subequations}
\begin{align}\label{OP}\
&\mathop {\max }\limits_{\left\{ {{\bf{w}}\left( t \right), {\bf{x}}, {\bf{f}}  } \right\}} {\rm{ }}{R_b}\left( t \right)\\
&{s.t.} \quad {\left\| {\bf{w}}\left( t \right) \right\|^2} \le {P_{max}},\label{power_constraintop}\\
&\quad \quad {x_1} = 0,{x_M} \le {D_{\max }}, \label{distance_constraint1op}\\
&\quad \quad {x_m} - {x_{m - 1}} \ge {D_{min}}, \forall m, \label{distance_constrai1nt2op}\\
&\quad \quad {f_C} \le {f_m} \le {f_C} + \Delta F, \forall m , \label{frequency_constraintop}\\
&\quad \quad {\left| {{{\bf{h}}_{{aw_k}}}{\bf{w}}\left( t \right)} \right|^2} \le \sigma _{{w_k}}^2\left( {{y^{ - 1}}\left( {2{\epsilon ^2}} \right) - 1} \right), k=1,...,K, \label{covert_constraintop}
\end{align}
\end{subequations}
where \eqref{power_constraintop} is the maximum power constraint and ${P_{max}}$ is the maximum transmit power. (15c) is the 1D linear range constraint which ensures that MFDA is movable within the predetermined region $\left[ {0,{D_{\max }}} \right]$ and ${{D_{\max }}}$ is the maximum movable distance of MFDA. Constraint \eqref{distance_constrai1nt2op} aims to avoid antenna coupling and ${D_{min}}$ is the minimum distance between any adjacent antennas at the MFDA transmitter. Besides, \eqref{frequency_constraintop} is the frequency constraint of MFDA, and \eqref{covert_constraintop} is the covert constraint for different Willies.

\section{Proposed Algorithm for Optimization Problem}

We observe that problem (15) contains three optimized variables which are tightly coupled making the problem challenging to solve. Usually, the AO algorithm is utilized to solve this problem \cite{8972400,9151964}. However, when there are multiple variables, the complexity of the AO algorithm increases significantly and the convergence accuracy decreases. To avoid these problems, we propose a low-complexity two-stage AO algorithm to solve problem (15).

\subsection{Optimizing ${\bf{w}}\left( t \right)$ with Given ${\bf{x}}$ and ${\bf{f}}$ }

According to \eqref{R_b}, problem (15) can be reformulated as
\begin{subequations}
\begin{align}\label{OP1}\
&\mathop {\max }\limits_{{{\bf{w}}\left( t \right)}} \quad{{\bf{w}}^\dag }\left( t \right){{\bf{H}}_{ab}}{\bf{w}}\left( t \right) \\
&{s.t.} \quad {\left\| {\bf{w}}\left( t \right) \right\|^2} \le {P_{max}},\label{power_constraint}\\
&\quad \quad {{\bf{w}}^\dag }\left( t \right){{\bf{H}}_{a{w_k}}}{\bf{w}}\left( t \right) \le \sigma _{{w_k}}^2\left( {{y^{ - 1}}\left( {2{\epsilon ^2}} \right) - 1} \right),k = 1,...,K, \label{covert_constraintop_1}
\end{align}
\end{subequations}
where ${{\bf{H}}_{ab}} = {\bf{h}}_{ab}^\dag {{\bf{h}}_{ab}}$ and ${{\bf{H}}_{a{w_k}}} = {\bf{h}}_{a{w_k}}^\dag {{\bf{h}}_{a{w_k}}}$. Note that ${{\bf{w}}^\dag }\left( t \right){{\bf{H}}_{ab}}{\bf{w}}\left( t \right) = \rm{tr}\left( {{{\bf{H}}_{ab}}{\bf{w}}\left( t \right){{\bf{w}}^\dag }\left( t \right)} \right)$. We define the matrix ${\bf{W}}\left( t \right) = {\bf{w}}\left( t \right){{\bf{w}}^\dag }\left( t \right)$ which implies that  ${\rm{rank}}\left( {\bf{W}} \right) = 1$, and the rank-1 constraint is non-convex which can be relaxed by the semi-definite relaxation (SDR) method. Then, problem (16) can be converted to
\begin{subequations}
\begin{align}\label{OP1_new1}\
&\mathop {\max }\limits_{{\bf{W}}\left( t \right)}\quad \rm{tr}\left( {{{\bf{H}}_{ab}}{\bf{W}}\left( t \right)} \right) \\
&{s.t.} \quad \rm{tr}\left( {{\bf{W}}\left( t \right)} \right) \le {P_{\max }},\label{power_constraint}\\
&\quad \quad {\rm{tr}}\left( {{{\bf{H}}_{{\rm{a}}{{\rm{w}}_k}}}{\bf{W}}\left( {\rm{t}} \right)} \right) \le \sigma _{{w_k}}^2\left( {{y^{ - 1}}\left( {2{\epsilon ^2}} \right) - 1} \right),k = 1,...,K. \label{covert_constraintop_1}
\end{align}
\end{subequations}
Problem (17) is a semi-definite program (SDP) problem which can be efficiently solved by convex solvers such as CVX \cite{2008CVX}. We observe that the optimal solution ${{\bf{W}}^{opt}}\left( t \right)$ may not satisfy the rank-1 constraint. Thus, we apply the Gaussian randomization method to recover ${{\bf{w}}^{opt}}\left( t \right)$ approximately \cite{8811733}.
\subsection{Optimizing ${\bf{x}}$ with Given ${\bf{w}}\left( t \right)$ and ${\bf{f}}$ }

It does not appear possible to derive an expression for the optimal ABV in closed form from problem (17). Hence, there is no simplified optimization object when optimizing the APV and AFV with given ABV. Fortunately, we note that FDA can make the channels of the legitimate and illegitimate users as orthogonal as possible to minimize the channel correlations by optimizing the AFV \cite{10283605,10097703,8078202,8458419}. Inspired by this observation, we assume that MFDA aims to minimize the sum of the inner product between the channels of Bob and Willies. In particular, the ideal case is ${{\bf{h}}_{ab}} \perp {{\bf{h}}_{aw_k}}$, $k=1,..,K$. Then, the optimization problem in this case can be formulated as
\begin{subequations}
\begin{align}\label{OP2}\
&\mathop {\min }\limits_{ {{\bf{x}} } } {\rm{ }} \sum\limits_{k = 1}^K {{{\left| {\left\langle {{{\bf{h}}_{ab}},{{\bf{h}}_{a{w_k}}}} \right\rangle } \right|}^2}}  \\
&{s.t.} \quad {x_1} = 0,{x_M} \le {D_{\max }}, \label{distance_constraint1op}\\
&\quad \quad {x_m} - {x_{m - 1}} \ge {D_{min}}, \forall m. \label{distance_constraint2op}
\end{align}
\end{subequations}
According to \eqref{h_ab} and \eqref{h_ae}, \eqref{OP2} can be written as
\begin{small}
 \begin{align}\label{habwithhae}\
&\sum\limits_{k = 1}^K {{{\left| {\left\langle {{{\bf{h}}_{ab}},{{\bf{h}}_{a{w_k}}}} \right\rangle } \right|}^2}}  \nonumber\\
 &= \sum\limits_{k = 1}^K {{{\left| {\sum\limits_{m = 1}^M {{e^{j2\pi {f_m}\left( {{x_m}\frac{{\sin {\theta _{{w_k}}} - \sin {\theta _b}}}{c} + \frac{{{r_b} - {r_{{w_k}}}}}{c}} \right)}}} } \right|}^2}}  \nonumber\\
  &= \sum\limits_{k = 1}^K {\sum\limits_{m = 1}^M {\sum\limits_{n = 1}^M {{e^{j2\pi \left( {\left( {{f_m}{x_m} - {f_n}{x_n}} \right)\frac{{\sin {\theta _{{w_k}}} - \sin {\theta _b}}}{c} + \left( {{f_m} - {f_n}} \right)\frac{{{r_b} - {r_{{w_k}}}}}{c}} \right)}}} } }
 \nonumber\\
 &= KM + \sum\limits_{k = 1}^K {\sum\limits_{m = 1}^M {\sum\limits_{\scriptstyle{n=1}\hfill\atop
\scriptstyle{n \ne  m}\hfill}^M \cos \left[ {2\pi \left( \begin{array}{l}
\left( {{f_m}{x_m} - {f_n}{x_n}} \right)\\
 \times \frac{{\sin {\theta _{{w_k}}} - \sin {\theta _b}}}{c}\\
 + \left( {{f_m} - {f_n}} \right)\frac{{{r_b} - {r_{{w_k}}}}}{c}
\end{array} \right)} \right] } } .
\end{align}
\end{small}
Next, problem (18) can be reformulated as
\begin{subequations}
\begin{align}\label{OP3}\
&\mathop {\min }\limits_{ {{\bf{x}}}} {\sum\limits_{k = 1}^K {\sum\limits_{m = 1}^M {\sum\limits_{\scriptstyle{n=1}\hfill\atop
\scriptstyle{n \ne  m}\hfill}^M \cos \left[ {2\pi \left( \begin{array}{l}
\left( {{f_m}{x_m} - {f_n}{x_n}} \right)\\
 \times \frac{{\sin {\theta _{{w_k}}} - \sin {\theta _b}}}{c}\\
 + \left( {{f_m} - {f_n}} \right)\frac{{{r_b} - {r_{{w_k}}}}}{c}
\end{array} \right)} \right]} } }\\
&{s.t.} \quad {x_1} = 0,{x_M} \le {D_{\max }}, \label{distance_constraint1}\\
&\quad \quad {x_m} - {x_{m - 1}} \ge {D_{min}}, \forall m. \label{distance_constraint2}
\end{align}
\end{subequations}
Problem (20) is clearly non-convex. Thus, we resort to the BSUM method to solve it. BSUM solves the problem by formulating a new convex function to approach the optimization target. Specifically, in the $s$-th iteration of BSUM, we first optimize ${{{x _m}}}$ where $m = s{\kern 1pt} {\kern 1pt} {\kern 1pt} {\kern 1pt} {\rm{mod}}{\kern 1pt} {\kern 1pt} {\kern 1pt} M$, and obtain the optimal position $x_m^{opt}$. Then, we update ${{\bf{x}}^{s - 1}}$ with $x_m^{opt}$, i.e., ${{\bf{x}}^s} = {{\bf{x}}^{s - 1}}$. In the $s$-th iteration, the optimization problem can be written as
\begin{subequations}
\begin{align}\label{OPAO1}\
&\mathop {\min }\limits_{{x_m}} \sum\limits_{k = 1}^K {\sum\limits_{\scriptstyle{n=1}\hfill\atop
\scriptstyle{n \ne  m}\hfill}^M \cos \left[ {2\pi \left( \begin{array}{l}
\left( {{f_m}{x_m} - {f_n}{x_n}} \right)\frac{{\sin {\theta _{{w_k}}} - \sin {\theta _b}}}{c}\\
 + \left( {{f_m} - {f_n}} \right)\frac{{{r_b} - {r_{{w_k}}}}}{c}
\end{array} \right)} \right] }\\
&{s.t.} \quad {x_1} = 0,{x_M} \le {D_{\max }}, \label{distance_constraintkk1}\\
&\quad \quad {x_m} - {x_{m - 1}} \ge {D_{min}}, \forall m. \label{distance_constraintkk2}
\end{align}
\end{subequations}
Next, we define the non-convex function in \eqref{OPAO1} as
\begin{align}\label{oldobject}\
{y_{k,m,n}}\left( x \right) = \cos \left[ {2\pi \left( \begin{array}{l}
\left( {{f_m}{x} - {f_n}{x_n}} \right)\frac{{\sin {\theta _{{w_k}}} - \sin {\theta _b}}}{c}\\
 + \left( {{f_m} - {f_n}} \right)\frac{{{r_b} - {r_{{w_k}}}}}{c}
\end{array} \right)} \right].
\end{align}
Following the BSUM method, the objective function \eqref{oldobject} is approximated by the upper-bound quadratic function ${u_{k,m,n}}\left( {{x}} \right)$, which is defined by
\begin{align}\label{newobject}\
{u_{k,m,n}}\left( x \right) = {A_{k,m,n}}{\left( {x - {B_{k,m,n}}} \right)^2} + {C_{k,m,n}},
\end{align}
where ${A_{k,m,n}} \in \mathbb{R}$, ${A_{k,m,n}} >0 $, ${{B_{k,m,n}}}\in \mathbb{R}$ and ${C_{k,m,n}}\in \mathbb{R}$ are the parameters of the new quadratic function. For a given point ${x_m^{s - 1}}$ in \eqref{oldobject}, the approximate function \eqref{newobject} should satisfy the following constraints:
\begin{align}\label{newobject_constraints}\
\left\{ {\begin{array}{*{20}{l}}
{{u_{k,m,n}}\left( {x_m^{s - 1};x_n^{s - 1}} \right) = {y_{k,m,n}}\left( {x_m^{s - 1};x_n^{s - 1}} \right)}\\
{u_{k,m,n}^{'}\left( {x_m^{s - 1};x_n^{s - 1}} \right) = y_{k,m,n}^{'}\left( {x_m^{s - 1};x_n^{s - 1}} \right)}\\
{{u_{k,m,n}}\left( {{B_{k,m,n}};x_n^{s - 1}} \right) \in \left\{ {1, - 1} \right\}}\\
 - 1 \le {y_{k,m,n}}\left( {{B_{k,m,n}};x_n^{s - 1}} \right) \le 1,
\end{array}} \right.
\end{align}
where $y_{k,m,n}^{'} \left( {x_m^{s - 1};x_n^{s - 1}} \right)=  - 2\pi {f_m}\frac{{\sin {\theta _{{w_k}}} - \sin {\theta _b}}}{c}  \times  \\ \sin \left[ {2\pi \left( {\left( {{f_m}x_m^{s - 1} - {f_n}{x_n}} \right)\frac{{\sin {\theta _{{w_k}}} - \sin {\theta _b}}}{c} + \left( {{f_m} - {f_n}} \right)\frac{{{r_b} - {r_{{w_k}}}}}{c}} \right)} \right]$. Next, we can obtain the parameters of the new quadratic function in \eqref{newobject} by utilizing Lemma 1.

{\emph{\textbf{Lemma 1:}}}
If $y_{k,m,n}^{'} \left( {x_m^{s - 1};x_n^{s - 1}} \right) \ne 0$, we have \eqref{nle0kxidelta}.
\begin{figure*}[ht]
\begin{align}\label{nle0kxidelta}\
\left\{ {\begin{array}{*{20}{l}}
{A_{k,m,n}} = \frac{{ - \pi {f_m}\frac{{\sin {\theta _{{w_k}}} - \sin {\theta _b}}}{c}}}{{x_m^{s - 1} - {B_{k,m,n}}}}\sin \left[ {2\pi \left( {\left( {{f_m}x_m^{s - 1} - {f_n}x_n^{s - 1}} \right)\frac{{\sin {\theta _{{w_k}}} - \sin {\theta _b}}}{c} + \left( {{f_m} - {f_n}} \right)\frac{{{r_b} - {r_{{w_k}}}}}{c}} \right)} \right]\\
{{B_{k,m,n}} = \frac{c}{{{f_m}\left( {\sin {\theta _{{w_k}}} - \sin {\theta _b}} \right)}}\left[ {\frac{{1 + 2\left( {\left\lceil {\left( {{f_m}x_m^{s - 1} - {f_n}x_n^{s - 1}} \right)\frac{{\sin {\theta _{{w_k}}} - \sin {\theta _b}}}{c} + \left( {{f_m} - {f_n}} \right)\frac{{{r_b} - {r_{{w_k}}}}}{c}} \right\rceil  - 1} \right)}}{2} - \left( {{f_m} - {f_n}} \right)\frac{{{r_b} - {r_{{w_k}}}}}{c}} \right] + \frac{{{f_n}x_n^{s - 1}}}{{{f_m}}}}\\
{C_{k,m,n}} = \cos \left[ {2\pi \left( {\left( {{f_m}x_m^{s - 1} - {f_n}x_n^{s - 1}} \right)\frac{{\sin {\theta _{{w_k}}} - \sin {\theta _b}}}{c} + \left( {{f_m} - {f_n}} \right)\frac{{{r_b} - {r_{{w_k}}}}}{c}} \right)} \right] - {A_{k,m,n}}{\left( {x_m^{s - 1} - {B_{k,m,n}}} \right)^2}.
\end{array}} \right.
\end{align}
\end{figure*}
If $y_{k,m,n}^{'}\left( {x_m^{s - 1};x_n^{s - 1}} \right) = 0$, we have
\begin{align}\label{n0kxidelta}\
\left\{ {\begin{array}{*{20}{l}}
{{A_{k,m,n}} =  - 2{{\left( {\pi {f_m}\frac{{\sin {\theta _{{w_k}}} - \sin {\theta _b}}}{c}} \right)}^2}{C_{k,m,n}}}\\
{{B_{k,m,n}} = x_m^{s - 1}}\\
{{C_{k,m,n}} = {y_{k,m,n}}\left( {x_m^{s - 1};x_n^{s - 1}} \right) \in \left\{ { - 1,1} \right\}.}
\end{array}} \right.
\end{align}
The proof of Lemma 1 is presented in Appendix A. Therefore problem (21a) can be replaced by
\begin{align}\label{OPAO1_122}\
\mathop {\min }\limits_{{x_m}} \sum\limits_{k = 1}^K {\sum\limits_{\scriptstyle{n=1}\hfill\atop
\scriptstyle{n \ne  m}\hfill}^M {{A_{k,m,n}}{{\left( {{x_m} - {B_{k,m,n}}} \right)}^2} + {C_{k,m,n}}} }.
\end{align}
It is clear that the optimal solution of problem (21) is given by
\begin{align}\label{optx}\
x^{opt} = \frac{{\sum\limits_{k = 1}^K {\sum\limits_{\scriptstyle{n=1}\hfill\atop
\scriptstyle{n \ne  m}\hfill}^M {{A_{k,m,n}}{B_{k,m,n}}} } }}{{\sum\limits_{k = 1}^K {\sum\limits_{\scriptstyle{n=1}\hfill\atop
\scriptstyle{n \ne  m}\hfill}^M {{A_{k,m,n}}} } }}.
\end{align}
According to the constraints \eqref{distance_constraintkk1} and \eqref{distance_constraintkk2}, the optimal position of the $m$-th antenna is given by
 \begin{align}\label{opt_xm}\
x_m^{opt} = \left\{ {\begin{array}{*{20}{l}}
{x_{m - 1}^s + {D_{\min }}}&{{x^{opt}} \le x_{m - 1}^s + {D_{\min }}}\\
{{x^{opt}}}&{x_{m - 1}^s + {D_{\min }} < {x^{opt}} \le {D_{upper}}}\\
{{D_{upper}}}&{{D_{upper}} < {x^{opt}}},
\end{array}} \right.
\end{align}
where ${D_{upper}} = {D_{\max }} - \left( {M - m} \right){D_{\min }}$. The convergence of BSUM is verified in Fig. 2.

\subsection{Optimizing ${\bf{f}}$ with Given ${\bf{w}}\left( t \right)$ and ${\bf{x}}$ }

With given ${\bf{w}}\left( t \right)$ and ${\bf{x}}$, the optimization problem can be formulated as
\begin{subequations}
\begin{align}\label{OPAO2}\
& \mathop {\min }\limits_{\bf{f}} \sum\limits_{k = 1}^K {\sum\limits_{m = 1}^M {\sum\limits_{\scriptstyle{n=1}\hfill\atop
\scriptstyle{n \ne  m}\hfill}^M {\cos } \left[ {2\pi \left( {{\tau _{k,m}}{f_m} - {\tau _{k,n}}{f_n}} \right)} \right]} } \\
&{s.t.} \quad {f_C} \le {f_m} \le {f_C} + \Delta F, \forall m, \label{frequency_constraint}
\end{align}
\end{subequations}
where ${\tau _{k,m}} = {x_m}\frac{{\sin {\theta _{{w_k}}} - \sin {\theta _b}}}{c} + \frac{{{r_b} - {r_{{w_k}}}}}{c}$. Problem (30) is non-convex which can be solved by the BSUM method developed in the previous subsection. Thus, we define the non-convex function in \eqref{OPAO2} as
\begin{align}\label{oldobject_f}\
{y_{k,m,n}}\left( f \right) = \cos \left[ {2\pi \left( {{\tau _{k,m}}{f} - {\tau _{k,n}}{f_n}} \right)} \right].
\end{align}
The new quadratic function is the same as \eqref{newobject}. Then, the parameters of the new quadratic function can be obtained by utilizing Lemma 2.

{\emph{\textbf{Lemma 2:}}}
If $y_{k,m,n}^{'} \left( {f_m^{s - 1};f_n^{s - 1}} \right) \ne 0$ and ${r_b} \ge {r_{{w_k}}}$, we have
\begin{align}\label{nle0kxidelta_f}\
\left\{ {\begin{array}{*{20}{l}}
{A_{k,m,n}} &= \frac{{ - \pi {\tau _{k,m}}\sin \left[ {2\pi \left( {f_m^{s - 1}{\tau _{k,m}} - f_n^{s - 1}{\tau _{k,n}}} \right)} \right]}}{{f_m^{s - 1} - {B_{k,m,n}}}}\\
{B_{k,m,n}} &= \frac{{1 + 2\left\lfloor {\left( {f_m^{s - 1}{\tau _{k,m}} - f_n^{s - 1}{\tau _{k,n}}} \right)} \right\rfloor }}{{2{\tau _{k,m}}}} + \frac{{f_n^{s - 1}{\tau _{k,n}}}}{{{\tau _{k,m}}}}\\
{C_{k,m,n}} &= \cos \left[ {2\pi \left( {f_m^{s - 1}{\tau _{k,m}} - f_n^{s - 1}{\tau _{k,n}}} \right)} \right]\\
 &- {A_{k,m,n}}{\left( {f_m^{s - 1} - {B_{k,m,n}}} \right)^2}.
\end{array}} \right.
\end{align}
If $y_{k,m,n}^{'} \left( {f_m^{s - 1};f_n^{s - 1}} \right) \ne 0$ and ${r_b} < {r_{{w_k}}}$, we have
\begin{align}\label{nle0kxidelta_f1}\
\left\{ {\begin{array}{*{20}{l}}
{A_{k,m,n}} &= \frac{{ - \pi {\tau _{k,m}}\sin \left[ {2\pi \left( {f_m^{s - 1}{\tau _{k,m}} - f_n^{s - 1}{\tau _{k,n}}} \right)} \right]}}{{f_m^{s - 1} - {B_{k,m,n}}}}\\
{B_{k,m,n}} &=  - \frac{{1 + 2\left\lfloor {\left( {f_n^{s - 1}{\tau _{k,n}} - f_m^{s - 1}{\tau _{k,m}}} \right)} \right\rfloor }}{{2{\tau _{k,m}}}} + \frac{{f_n^{s - 1}{\tau _{k,n}}}}{{{\tau _{k,m}}}}\\
{C_{k,m,n}} &= \cos \left[ {2\pi \left( {f_m^{s - 1}{\tau _{k,m}} - f_n^{s - 1}{\tau _{k,n}}} \right)} \right]\\
 &- {A_{k,m,n}}{\left( {f_m^{s - 1} - {B_{k,m,n}}} \right)^2}.
\end{array}} \right.
\end{align}
If $y_{k,m,n}^{'}\left( {f_m^{s - 1};f_n^{s - 1}} \right) = 0$, we have
\begin{align}\label{n0kxidelta_f}\
\left\{ {\begin{array}{*{20}{l}}
{{A_{k,m,n}} = - 2{\pi ^2}\tau _{k,m}^2{C_{k,m,n}} }\\
{{B_{k,m,n}} = f_m^{s - 1}}\\
{{C_{k,m,n}} = {y_{k,m,n}}\left( {f_m^{s - 1};f_n^{s - 1}} \right) \in \left\{ { - 1,1} \right\}},
\end{array}} \right.
\end{align}
where $y_{k,m,n}^{'}\left( f \right) =  - 2\pi {\tau _{k,m}}\sin \left[ {2\pi \left( {{\tau _{k,m}}f - {\tau _{k,n}}{f_n}} \right)} \right]$. Due to the similar mathematical proof as Lemma 1, we omit the proof of Lemma 2 here for brevity. Then, the optimal frequency of the $m$-th antenna is given by
 \begin{align}\label{opt_fm}\
f_m^{opt} = \left\{ {\begin{array}{*{20}{l}}
{{f_C}}&{{f_{opt}} < {f_C}}\\
{{f^{opt}}}&{{f_C} \le {f^{opt}} < {f_C} + \Delta F}\\
{{f_C} + \Delta F}&{{f_C} + \Delta F \le {f^{opt}}},
\end{array}} \right.
\end{align}
where ${{f^{opt}}}$ can be obtained from \eqref{optx} with the parameters in Lemma 2.

\subsection{Overall Two-Stage AO Algorithm}
According to problems (20) and (30), it is clear that ABV cannot affect the optimization for APV or AFV. It means that ABV is not coupled with APV and AFV, but APV and AFV are coupled with each other. Thus, we propose a two-stage AO algorithm with low complexity. Specifically, the first stage aims to alternately optimize APV and AFV by utilizing the BSUM method which is summarized in Algorithm 1.
\begin{algorithm}
\caption{Optimizing ${\bf{x}}$ by BSUM}
\begin{algorithmic}[1]
\STATE {\bf{Initialization:}} Initializing ${\bf{f}}$ via Algorithm 1 .
\STATE {\bf{Set:}} $s=0$.
\STATE {{\bf{Repeat:}}\\
        a) $s=s+1$, and $m = s{\kern 1pt} {\kern 1pt} {\kern 1pt} {\kern 1pt} {\rm{mod}}{\kern 1pt} {\kern 1pt} {\kern 1pt} {\rm{M}}$.\\
        b) The new quadratic function ${u_{k,m,n}}\left( {{x}} \right)$ can be obtained by Lemma 1. Then, the optimal position of $m$-th antenna can be obtained by \eqref{opt_xm}.\\
        c) We update ${{\bf{x}}^{s-1}}$ with $x_m^{opt}$ and obtain ${{\bf{x}}^s}$.\\}
\STATE {{\bf{Until}}  $\left| \begin{array}{l}
\sum\limits_{k = 1}^K {{{\left| {\left\langle {{{\bf{h}}_{ab}}\left( {{{\bf{x}}^s},{\bf{f}}} \right),{{\bf{h}}_{a{w_k}}}\left( {{{\bf{x}}^s},{\bf{f}}} \right)} \right\rangle } \right|}^2}} \\
 - \sum\limits_{k = 1}^K {{{\left| {\left\langle {{{\bf{h}}_{ab}}\left( {{{\bf{x}}^{s - 1}},{\bf{f}}} \right),{{\bf{h}}_{a{w_k}}}\left( {{{\bf{x}}^{s - 1}},{\bf{f}}} \right)} \right\rangle } \right|}^2}}
\end{array} \right| \le \varsigma $ is satisfied where $\varsigma$ is a small convergence value.}
\end{algorithmic}
\end{algorithm}
The second stage is to obtain the optimal ABV with the optimized APV and AFV. The overall algorithm is summarized in Algorithm 2.
\begin{algorithm}
\caption{Overall two-stage AO algorithm}
\begin{algorithmic}[1]
\STATE {\bf{Initialization:}} Initializing ${{{\bf{w}}_{opt}}}$ via \eqref{power_constraintop}.
\STATE {\bf{Set:}} $i=0$.
\STATE {{\bf{Repeat:}}\\
        a) $i=i+1$.\\
        b) Optimizing  ${\bf{x}}$ and ${\bf{f}}$ alternately by BSUM.\\
        c) Obtaining  ${{{\bf{w}}_{opt}}}$ with the optimized ${{\bf{x}}^s}$ and ${{\bf{f}}^s}$ by solving problem (17).\\}
\STATE {{\bf{Until}} $\left| {{R_b}\left( {{{\bf{w}}_{opt}},{{\bf{x}}^s},{{\bf{f}}^s}} \right) - {R_b}\left( {{{\bf{w}}_{opt}},{{\bf{x}}^{s - 1}},{{\bf{f}}^{s - 1}}} \right)} \right| \le \varsigma $ is satisfied.}
\end{algorithmic}
\end{algorithm}

 The complexity of Algorithm 1 is ${\cal{O}}\left( {{I_{BSUM}}} \right)$ where ${{I_{BSUM}}}$ denotes the number of iterations required for BSUM convergence. The overall complexity of Algorithm 2 is ${\cal{O}}\left( {{I_{AO}}\left( {{I_{BSUM}} + {I_{BSUM}}} \right)} + {M^{3.5}}\log \left( {\frac{1}{\varsigma }} \right) \right)$ where ${{I_{AO}}}$ denotes the number of iterations required for AO algorithm convergence.

\subsection{Special Case: Perfect Covert Communication}

Based on the perfect CSI at Willies, the perfect covertness may be achieved which means that no signal power can be detected by Willies, i.e., ${{\bf{w}}^\dag }\left( t \right){{\bf{H}}_{a{w_k}}}{\bf{w}}\left( t \right) = 0,k = 1,...,K,$. The optimization problem in this case can be formulated as
\begin{subequations}
\begin{align}\label{OP_specialcase1}\
&\mathop {\max }\limits_{\left\{ {{\bf{w}}\left( t \right), {\bf{x}}, {\bf{f}}  } \right\}} {\rm{ }}{{\bf{w}}^\dag }\left( t \right){{\bf{H}}_{ab}}{\bf{w}}\left( t \right) \\
&{s.t.} \quad {\left\| {\bf{w}}\left( t \right) \right\|^2} \le {P_{max}},\label{power_constraintop_special}\\
&\quad \quad {x_1} = 0,{x_M} \le {D_{\max }}, \label{distance_constraint1opspecial}\\
&\quad \quad {x_m} - {x_{m - 1}} \ge {D_{min}}, \forall m, \label{distance_constraint2opspecial}\\
&\quad \quad {f_C} \le {f_m} \le {f_C} + \Delta F, \forall m , \label{frequency_constraintopspecial}\\
&\quad \quad {{\bf{w}}^\dag }\left( t \right){{\bf{H}}_{a{w_k}}}{\bf{w}}\left( t \right) =0, k=1,...,K. \label{covert_constraintopspecial}
\end{align}
\end{subequations}
Problem (36) can be solved by the two-stage AO algorithm proposed in the previous subsection. With given  ${\bf{f}}$ and ${\bf{x}}$, problem (36) can be rewritten as
\begin{subequations}
\begin{align}\label{OP_specialcase2}\
&\mathop {\max }\limits_{ {{\bf{w}}\left( t \right) } } {\rm{ }}{{\bf{w}}^\dag }\left( t \right){{\bf{H}}_{ab}}{\bf{w}}\left( t \right) \\
&{s.t.} \quad {\left\| {\bf{w}}\left( t \right) \right\|^2} \le {P_{max}},\label{power_constraintop_specia2}\\
&\quad \quad {{\bf{w}}^\dag }\left( t \right){{\bf{H}}_{a{w_k}}}{\bf{w}}\left( t \right) =0, k=1,...,K. \label{covert_constraintopspecia2}
\end{align}
\end{subequations}
The optimal solution to problem (37) is given as \cite{9122034}
\begin{align}\label{w_opt_special1}\
{{\bf{w}}_{opt}}\left( t \right) = \sqrt {{P_{\max }}} \frac{{\left( {{I_M} - {\bf{H}}_{{\bf{aw}}}^\dag {{\left( {{{\bf{H}}_{{\bf{aw}}}}{\bf{H}}_{{\bf{aw}}}^\dag } \right)}^{ - 1}}{{\bf{H}}_{{\bf{aw}}}}} \right){\bf{h}}_{ab}^\dag }}{{\left\| {\left( {{I_M} - {\bf{H}}_{{\bf{aw}}}^\dag {{\left( {{{\bf{H}}_{{\bf{aw}}}}{\bf{H}}_{{\bf{aw}}}^\dag } \right)}^{ - 1}}{{\bf{H}}_{{\bf{aw}}}}} \right){\bf{h}}_{ab}^\dag } \right\|}},
\end{align}
where ${{\bf{H}}_{{\bf{aw}}}} = \left[ {{{\bf{h}}_{a{w_1}}};...;{{\bf{h}}_{a{w_K}}}} \right]$. The proof is presented in Appendix B. Then, with given ${{\bf{w}}_{opt}}\left( t \right)$, the optimal ${\bf{f}}$ and ${\bf{x}}$ can be obtained using BSUM. The optimization process in this case is the same as that in the previous section and we omit it here for brevity.

\section{Imperfect CSI at Eve}

In this section, we consider a more practical scenario where Alice has imperfect CSIs knowledge about Willies. According to \eqref{h_ab} and \eqref{h_ae}, we conclude that the imperfect CSIs at Willies are mainly caused by an estimation error in Willies' ranges and angles. Without loss of generality, we assume that the exact coordinates of Willies are unknown, but the uncertain regions for Willies are known. For the $k$-th Willie, the uncertain region is denoted as ${{\cal{R}}_{{w_k}}} \buildrel \Delta \over = \left[ {{r_{{w_k}}} - \Delta {r_{{w_k}}},{r_{{w_k}}} + \Delta {r_{{w_k}}}} \right]$ and ${{\cal{A}}_{{w_k}}} \buildrel \Delta \over = \left[ {{\theta _{{w_k}}} - \Delta {\theta _{{w_k}}},{\theta _{{w_k}}} + \Delta {\theta _{{w_k}}}} \right]$, where ${{\Delta r}_{{w_k}}}$ and ${\Delta {\theta _{{w_k}}}}$ are the estimation errors in range and angle, respectively. Since ${{\cal{R}}_{{w_k}}}$ and ${{\cal{A}}_{{w_k}}}$ are continuous sets, we propose to uniformly generate some discrete samples over the sets which are given as ${r_{k,l}} = {r_{{w_k}}} - \Delta {r_{{w_k}}} + \left( {l - 1} \right)\frac{{2\Delta {r_{{w_k}}}}}{{L - 1}},l \in \left\{ {1,2,...,L} \right\}$ and ${\theta _{k,l}} = {\theta _{{w_k}}} - \Delta {\theta _{{w_k}}} + \left( {l - 1} \right)\frac{{2\Delta {\theta _{{w_k}}}}}{{L - 1}}$ where $L$ denotes the number of the samples. The optimization problem can be formulated as
\begin{subequations}
\begin{align}\label{OP_imperfect1}\
&\mathop {\max }\limits_{\left\{ {{\bf{w}}\left( t \right), {\bf{x}}, {\bf{f}}  } \right\}} {\rm{ }}{{R_b}\left( t \right)}\\
&{s.t.} \quad {\left\| {\bf{w}}\left( t \right) \right\|^2} \le {P_{max}},\label{power_constraintop_imperfect1}\\
&\quad \quad {x_1} = 0,{x_M} \le {D_{\max }}, \label{distance_constraint1op_imperfect1}\\
&\quad \quad {x_m} - {x_{m - 1}} \ge {D_0}, \forall m, \label{distance_constraint2op_imperfect1}\\
&\quad \quad {f_C} \le {f_m} \le {f_C} + \Delta F, \forall m .\label{frequency_constraintop_imperfect1}\\
&\quad \quad \begin{array}{l}
{\left| {{{\bf{h}}_{a{w_k}}}\left( {{r_{k,l}},{\theta _{k,l}}} \right){\bf{w}}\left( t \right)} \right|^2} \le \sigma _{{w_k}}^2\left( {{y^{ - 1}}\left( {2{\epsilon ^2}} \right) - 1} \right),\\
k = 1,...,K,l = 1,...,L.
\end{array}\label{covert_constraintop_new}
\end{align}
\end{subequations}
Problem (39) is a non-convex function with coupled variables which can be solved by our proposed two-stage AO algorithm.

\subsection{Optimizing ${\bf{w}}\left( t \right)$ with Given ${\bf{x}}$ and ${\bf{f}}$ }
 For given APV and AFV, problem (39) is converted to

 \begin{subequations}
\begin{align}\label{OP1_new1_imperfecty}\
&\mathop {\max }\limits_{{\bf{W}}\left( t \right)}\quad \rm{tr}\left( {{{\bf{H}}_{ab}}{\bf{W}}\left( t \right)} \right) \\
&{s.t.} \quad \rm{tr}\left( {{\bf{W}}\left( t \right)} \right) \le {P_{\max }},\label{power_constraint_imperfecty}\\
&\quad \quad\begin{array}{l}
{\rm{tr}}\left( {{{\bf{H}}_{{\rm{a}}{{\rm{w}}_k}}}{\bf{W}}\left( {\rm{t}} \right)} \right) \le \sigma _{{w_k}}^2\left( {{y^{ - 1}}\left( {2{\epsilon ^2}} \right) - 1} \right),\\
k = 1,...,K,l = 1,...,L.
\end{array} \label{covert_constraintop_1_imperfecty}
\end{align}
\end{subequations}
Problem (40) is a semi-definite program (SDP) problem which is efficiently solved by CVX. If the rank-1 constraint is not satisfied, the Gussian randomization method is applied to recover ${{\bf{w}}^{opt}}$ approximately.

\subsection{Optimizing ${\bf{x}}$ with Given ${\bf{w}}\left( t \right)$ and ${\bf{f}}$. }

Considering the coordinates uncertainty at Willies, we aim to minimize the sum of the channel correlations between the Alice$\to $Bob link and all Alice$ \to$Willies links with Willies located at the different samples in the uncertainty region. The problem is formulated as
\begin{subequations}
\begin{align}\label{OP2_imperfect}\
&\mathop {\min }\limits_{ {{\bf{x}} } } {\rm{ }}\sum\limits_{k = 1}^K {\sum\limits_{i = 1}^L {\sum\limits_{j = 1}^L {{{\left| {\left\langle {{{\bf{h}}_{ab}},{{\bf{h}}_{a{w_k}}}\left( {{r_{k,i}},{\theta _{k,j}}} \right)} \right\rangle } \right|}^2}} } }  \\
&{s.t.} \quad {x_1} = 0,{x_M} \le {D_{\max }},  \label{distance_constraint1}\\
&\quad \quad {x_m} - {x_{m - 1}} \ge {D_{min}}, \forall m.  \label{distance_constraint2}
\end{align}
\end{subequations}
According to \eqref{habwithhae}, problem (41) is rewritten as
\begin{small}
\begin{subequations}
\begin{align}\label{OP2_imperfect11}\
&\mathop {\min }\limits_{ {{\bf{x}} } } {\rm{ }}\sum\limits_{k = 1}^K {\sum\limits_{i = 1}^L {\sum\limits_{j = 1}^L {\sum\limits_{m = 1}^M {\sum\limits_{\scriptstyle{n=1}\hfill\atop
\scriptstyle{n \ne  m}\hfill}^M  {\cos } \left[ {2\pi \left( {\begin{array}{*{20}{l}}
{\left( {{f_m}{x_m} - {f_n}{x_n}} \right)}\\
{ \times \frac{{\sin {\theta _{k,j}} - \sin {\theta _b}}}{c}}\\
{ + \left( {{f_m} - {f_n}} \right)\frac{{{r_b} - {r_{k,i}}}}{c}}
\end{array}} \right)} \right]} } } }  \\
&{s.t.} \quad {x_1} = 0,{x_M} \le {D_{\max }},  \label{distance_constraint1_imperfect}\\
&\quad \quad {x_m} - {x_{m - 1}} \ge {D_{min}}, \forall m.  \label{distance_constraint2_impferect}
\end{align}
\end{subequations}
\end{small}
Problem (42) is non-convex and is solved by BSUM. According to \eqref{OPAO1_122}, \eqref{OP2_imperfect11} is replaced by
\begin{subequations}
\begin{align}\label{OPAO1_122_imperfect}\
\mathop {\min }\limits_{{x_m}}\sum\limits_{k = 1}^K {\sum\limits_{i = 1}^L {\sum\limits_{j = 1}^L {\sum\limits_{\scriptstyle{n=1}\hfill\atop
\scriptstyle{n \ne  m}\hfill}^M {{A_{k,m,n}}{{\left( {{x_m} - {B_{k,m,n}}} \right)}^2} + {C_{k,m,n}}} } } },
\end{align}
\end{subequations}
where ${{A_{k,m,n}}}$, ${{B_{k,m,n}}}$ and ${{C_{k,m,n}}}$ are obtained using Lemma 1. Then, the optimal solution of \eqref{OPAO1_122_imperfect} is given by
\begin{align}\label{optx_imperfect}\
{x^{opt}} = \frac{{\sum\limits_{k = 1}^K {\sum\limits_{i = 1}^L {\sum\limits_{j = 1}^L {\sum\limits_{\scriptstyle{n=1}\hfill\atop
\scriptstyle{n \ne  m}\hfill}^M {{A_{k,m,n}}{B_{k,m,n}}} } } } }}{{\sum\limits_{i = 1}^L {\sum\limits_{j = 1}^L {\sum\limits_{\scriptstyle{n=1}\hfill\atop
\scriptstyle{n \ne  m}\hfill}^M {{A_{k,m,n}}} } } }}.
\end{align}
Considering the constraints \eqref{distance_constraint1_imperfect} and \eqref{distance_constraint2_impferect}, the optimal position of the $m$-th antenna is given in \eqref{opt_xm}.

\subsection{Optimizing ${\bf{f}}$ with Given ${\bf{w}}\left( t \right)$ and ${\bf{x}}$. }
Similar to problem (41), the optimization problem in this case is formulated as
\begin{subequations}
\begin{align}\label{OP2_imperfect_f}\
&\mathop {\min }\limits_{ {{\bf{f}} } } {\rm{ }}\sum\limits_{k = 1}^K {\sum\limits_{i = 1}^L {\sum\limits_{j = 1}^L {{{\left| {\left\langle {{{\bf{h}}_{ab}},{{\bf{h}}_{a{w_k}}}\left( {{r_{k,i}},{\theta _{k,j}}} \right)} \right\rangle } \right|}^2}} } }  \\
&{s.t.} \quad {f_C} \le {f_m} \le {f_C} + \Delta F, \forall m. \label{distance_constraint2_f_imprefct}
\end{align}
\end{subequations}
We propose to utilize the BSUM method to solve this problem. According to \eqref{OPAO1_122}, \eqref{OP2_imperfect_f} can be replaced by
\begin{subequations}
\begin{align}\label{OPAO1_122_imperfect_f}\
\mathop {\min }\limits_{{f_m}}\sum\limits_{k = 1}^K {\sum\limits_{i = 1}^L {\sum\limits_{j = 1}^L {\sum\limits_{\scriptstyle{n=1}\hfill\atop
\scriptstyle{n \ne  m}\hfill}^M {{A_{k,m,n}}{{\left( {{x_m} - {B_{k,m,n}}} \right)}^2} + {C_{k,m,n}}} } } },
\end{align}
\end{subequations}
where ${{A_{k,m,n}}}$, ${{B_{k,m,n}}}$ and ${{C_{k,m,n}}}$ are obtained using Lemma 2. Then, the optimal solution of \eqref{OPAO1_122_imperfect_f} is given as \eqref{optx_imperfect}. Considering the constraint \eqref{distance_constraint2_f_imprefct}, the optimal frequency of the $m$-th antenna is given as \eqref{opt_fm}.

{\emph{Remark 1:}}
According to \eqref{habwithhae}, it is clear that ${{{\left| {\left\langle {{{\bf{h}}_{ab}},{{\bf{h}}_{a{w_k}}}\left( {{r_{k,i}},{\theta _{k,j}}} \right)} \right\rangle } \right|}^2}}$ is not monotonic for ${r_{k,i}}$ and ${\theta _{k,j}}$, which means that we cannot directly find the worst-case coordinates for Willie in the estimated region $\left( {\cal{R}},{\cal{A}} \right)$. To address this issue, we utilize the discretization method to simplify the region to some samples which can effectively deal with the position uncertainty in FDA \cite{10097703}. It is worth mentioning that more samples can make the results accurate, but it also leads to higher complexity of the optimization.

\section{Numerical Results}

In this section, numerical results for MFDA-aided covert communication are presented. Unless otherwise specified, the maximum transmission power is ${P_{max}} = 10$dBm and the carrier frequency is ${f_C} = 10$GHz. The noise variances at Bob and Willies are $\sigma^2_{b}=-100$dBm and $\sigma _{{w_k}}^2 =  - 100$dBm, $k=1,...,K$. The position coordinate of Bob is $\left( {{r_b},{\theta _b}} \right) = \left( {1000{\rm{m}},{{30}^\circ }} \right)$. We consider four Willies, i.e., $K=4$, and the cases of low correlation channel and high correlation channel in this paper. In the case of low correlation channel, Willies are not close to Bob and located at $\left( {{r_{{w_1}}},{\theta _{{w_1}}}} \right) = \left( {1050{\rm{m}},{{10}^\circ }} \right)$, $\left( {{r_{{w_2}}},{\theta _{{w_2}}}} \right) = \left( {950{\rm{m}},{{20}^\circ }} \right)$, $\left( {{r_{{w_3}}},{\theta _{{w_3}}}} \right) = \left( {1100{\rm{m}},{{40}^\circ }} \right)$ and $\left( {{r_{{w_4}}},{\theta _{{w_4}}}} \right) = \left( {900{\rm{m}},{{50}^\circ }} \right)$, respectively. In the case of high correlation channel, Willies are close to Bob and located at $\left( {{r_{{w_1}}},{\theta _{{w_1}}}} \right) = \left( {1010{\rm{m}},{{26}^\circ }} \right)$, $\left( {{r_{{w_2}}},{\theta _{{w_2}}}} \right) = \left( {990{\rm{m}},{{28}^\circ }} \right)$, $\left( {{r_{{w_3}}},{\theta _{{w_3}}}} \right) = \left( {1020{\rm{m}},{{32}^\circ }} \right)$ and $\left( {{r_{{w_4}}},{\theta _{{w_4}}}} \right) = \left( {980{\rm{m}},{{34}^\circ }} \right)$, respectively. The minimum antenna separation distance is ${D_0} = \frac{\lambda }{2}$ where $\lambda  = \frac{c}{{{f_C}}}$ is the wavelength of MFDA. The movable antenna distance is ${D_{\max }} = 30\lambda $ and the frequency increment range is $\Delta F = 10$ MHz. Moreover, the path loss factor is given as \cite{8811733}
\begin{align}\label{LFS}\
{\rm{Lfs}}\left( r \right) = C{\left( {\frac{r}{R}} \right)^{ - \alpha }},
\end{align}
where $C=-30$dBm is the path loss at the reference distance $R=1{\rm{m}}$, $r$ is the individual link distance and $\alpha$ is the path loss exponent where ${\alpha _{ab}} = 2$ and ${\alpha _{ae}} = 3$.

For comparison, the covert rates of three benchmark strategies are also investigated as follows: 1) PA, i.e., only optimizing the ABV at Alice. 2) FDA, i.e., optimizing the ABV and AFV at Alice. 3) Perfect covertness, i.e., $ {{\bf{w}}^\dag }\left( t \right){{\bf{H}}_{a{w_k}}}{\bf{w}}\left( t \right) =0, k=1,...,K$. 4) Upper bound, i.e., $R_b = {\log}\left( {1 + \sigma _b^{ - 2}{\rm{Lf}}{{\rm{s}}^2}\left( {{r_b}} \right){P_{\max }}M} \right)$.

\begin{figure}[htbp]
	\centering
	\begin{minipage}{0.49\linewidth}
		\centering
		\subfigure[BSUM algorithm for APV]{\includegraphics[width=1.1\textwidth]{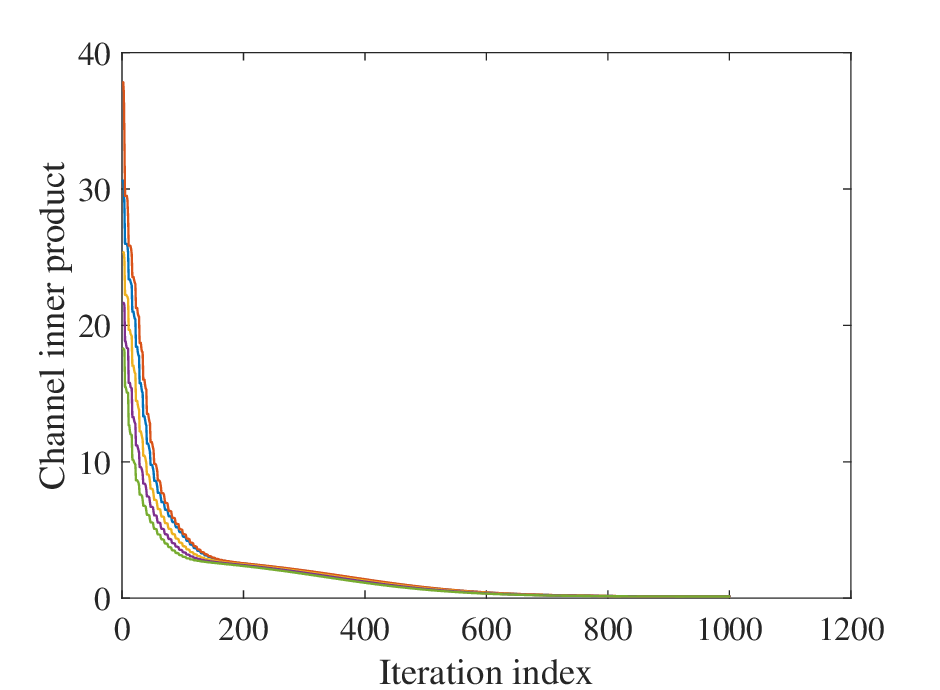}}
	\end{minipage}
	\begin{minipage}{0.49\linewidth}
		\centering
		\subfigure[BSUM algorithm for AFV]{\includegraphics[width=1.1\textwidth]{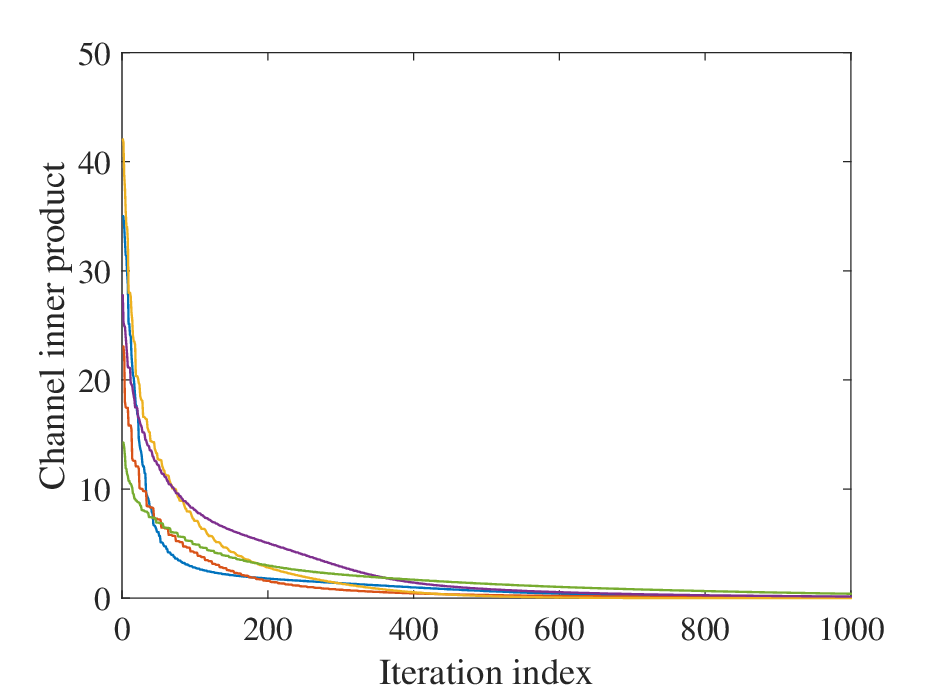}}
	\end{minipage}
\caption{Typical converging traces of the BSUM for optimizing APV and AFV with $M=10$.}
\label{fig2}
\end{figure}

Figure 2(a) demonstrates the convergence of the BSUM algorithm for optimizing APV where each trace starting from random initial APV and channel inner product is denoted as $\sum\limits_{k = 1}^K {{{\left| {\left\langle {{{\bf{h}}_{ab}},{{\bf{h}}_{a{w_k}}}} \right\rangle } \right|}^2}} $. The channel inner product is a metric for measuring the channel correlation between the legitimate channel and illegitimate channels. In particular, $\sum\limits_{k = 1}^K {{{\left| {\left\langle {{{\bf{h}}_{ab}},{{\bf{h}}_{a{w_k}}}} \right\rangle } \right|}^2}}  = 0$ means that ${{\bf{h}}_{ab}} \perp {{\bf{h}}_{aw_k}}$, $k=1,..,K$, which is the ideal case for optimizing the APV and AFV. We observe that BSUM takes about 800 iterations to converge and reach the minimum values. Moreover, the convergence behaviour for optimizing AFV is illustrated in Figure 7(b).

\begin{figure}[htbp]
	\centering
	\begin{minipage}{0.49\linewidth}
		\centering
		\subfigure[Low correlation channel]{\includegraphics[width=1.1\textwidth]{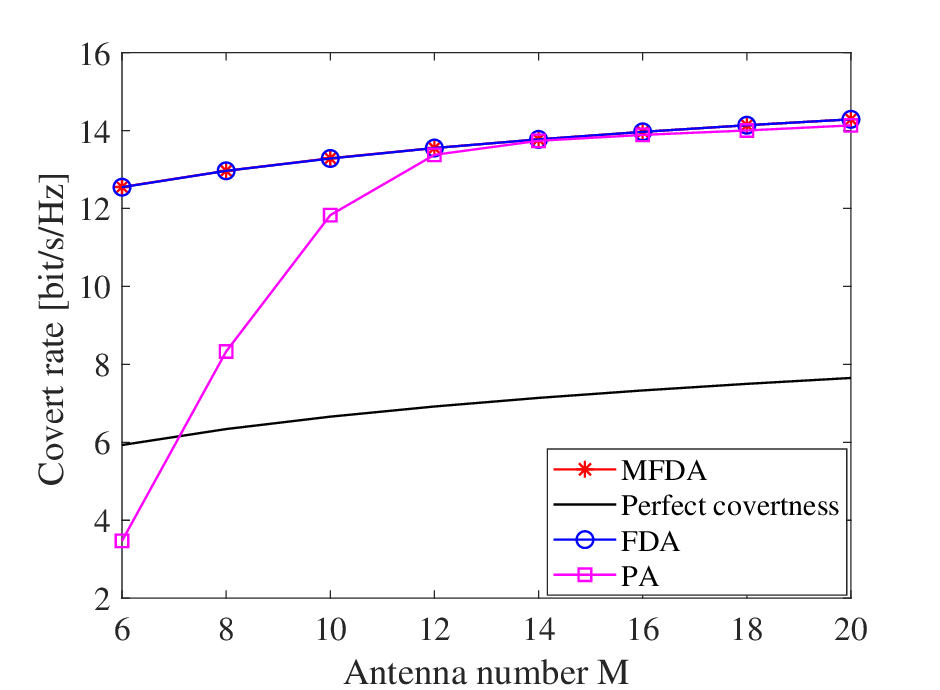}}
	\end{minipage}
	\begin{minipage}{0.49\linewidth}
		\centering
		\subfigure[High correlation channel]{\includegraphics[width=1.1\textwidth]{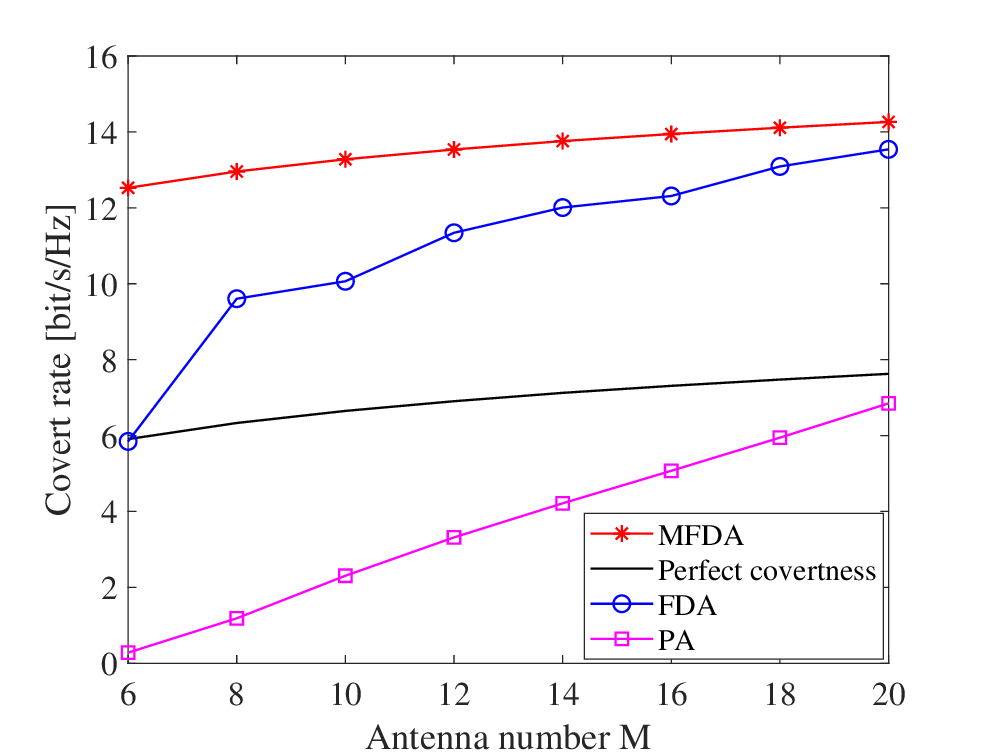}}
	\end{minipage}
\caption{Covert rate ${R_b}$ versus number of antennas $M$.}
\label{fig3}
\end{figure}

Figure 3 depicts the covert rate $R_b$ versus the number of antennas $M$ for different strategies. We observe that $R_b$ increases with $M$, because that brings more spatial DoFs. Besides, the covert rate of MFDA and FDA are larger than that of PA. This is due to the fact that PA only optimizes the ABV at Alice which results in a worse covert performance than other strategies as expected. In the case of the low correlation channel, the covert rate of MFDA is the same as that of FDA. However, in the case of the the high correlation channel, MFDA outperforms FDA. This is because FDA can only adjust the frequency, which is easily limited by the frequency constraint in a highly correlated channel scenario. In particular, MFDA can jointly adjust the frequency and position to achieve two-dimensional covertness and effectively handle different channel scenarios. Moreover, the perfect covertness strategy needs to ensure that Willies do not detect any signal power, thus cannot achieve larger covert rate due to the strict covert constraints.

\begin{figure}[htbp]
	\centering
	\begin{minipage}{0.49\linewidth}
		\centering
		\subfigure[Maximum distance]{\includegraphics[width=1.1\textwidth]{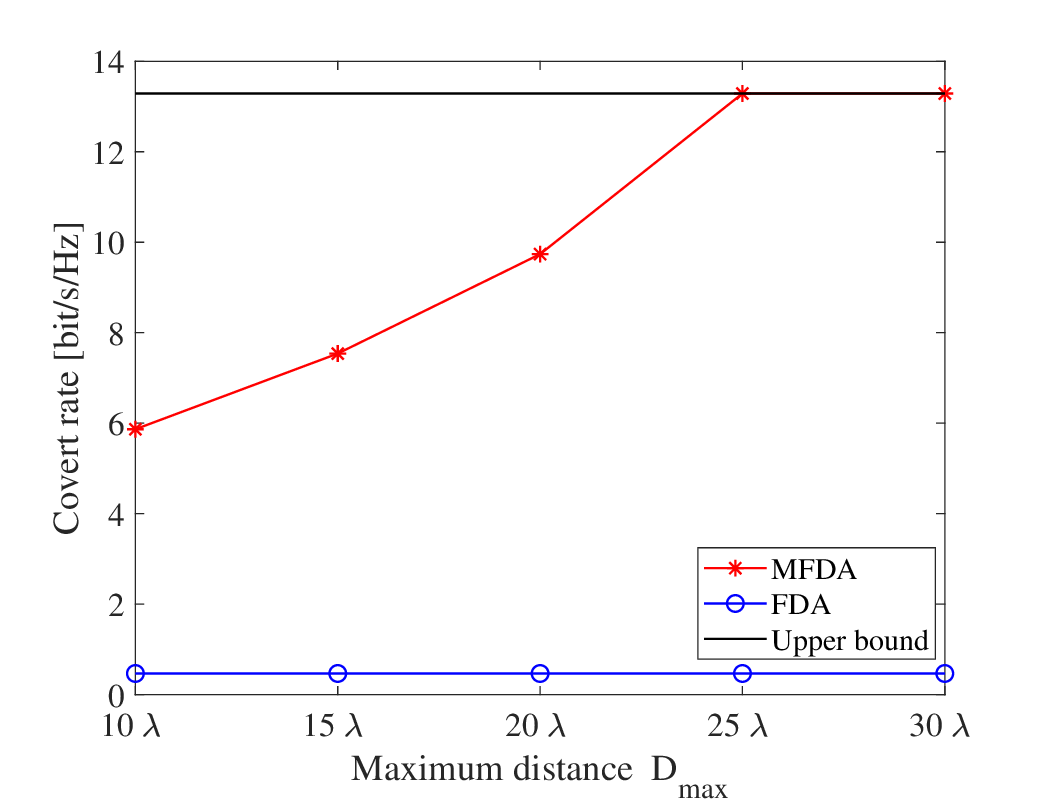}}
	\end{minipage}
	\begin{minipage}{0.49\linewidth}
		\centering
		\subfigure[Frequency increment range]{\includegraphics[width=1.1\textwidth]{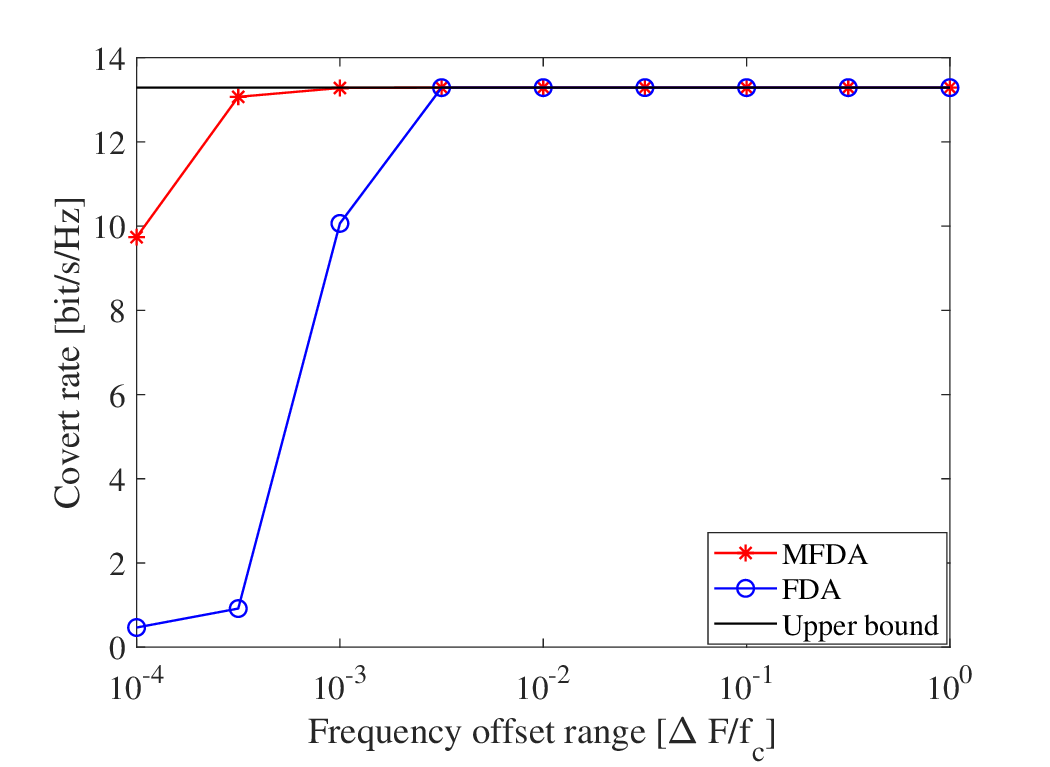}}
	\end{minipage}
\caption{Covert rate ${R_b}$ versus maximum distance ${D_{\max }}$ and frequency increment range $\Delta F/{f_c}$.}
\label{fig4}
\end{figure}

Figure 4 depicts the covert rate $R_b$ for different frequency increment range $\Delta F/{f_C}$ and movable antenna maximum distance ${D_{\max }}$. It is clear that larger ${D_{\max }}$ and $\Delta F$ can provide more selectivity to optimize APV and AFV which can improve the covert rate correspondingly. In practice, due to the device limitation or complex channel state, both APV and AFV will have constraints, such that relying on separately optimizing either one of them cannot achieve the best covert performance. Our proposed MFDA can jointly optimize APV and AFV to realize two-dimensional covertness which leads to a larger covert rate than FDA and PA. In particular, when the AFV constraint is strict, MFDA can optimize APV to further enhance covert performance instead of optimizing AFV.

\begin{figure}[htbp]
	\centering
	\begin{minipage}{0.49\linewidth}
		\centering
		\subfigure[Low correlation channel]{\includegraphics[width=1.1\textwidth]{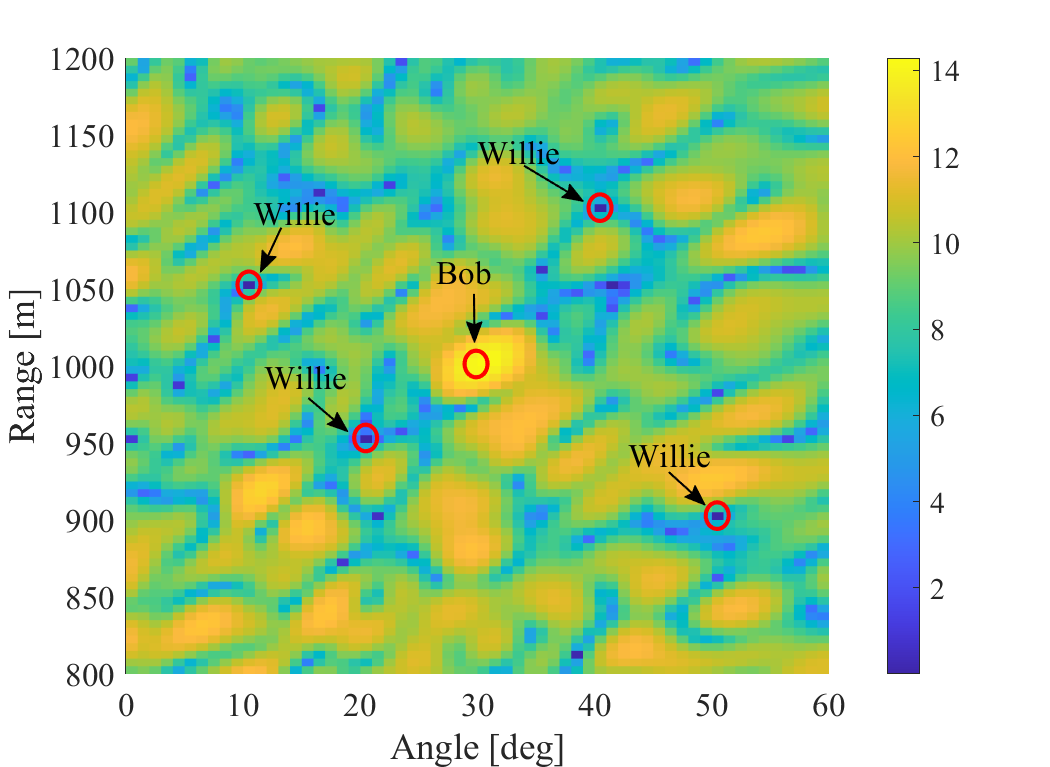}}
	\end{minipage}
	\begin{minipage}{0.49\linewidth}
		\centering
		\subfigure[High correlation channel]{\includegraphics[width=1.1\textwidth]{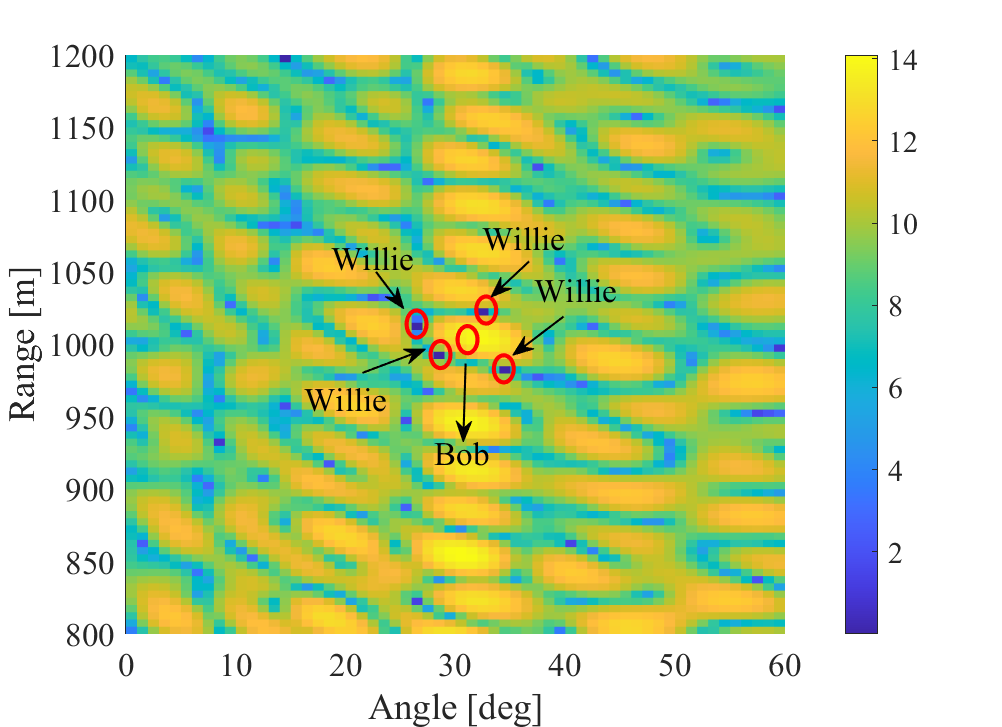}}
	\end{minipage}
\caption{Typical beampatterns for MFDA with $M=20$.}
\label{fig5}
\end{figure}

Figure 5 illustrates the beampattern with the optimal ABV, APV and AFV. We observe that MFDA can produce a ``dot-$shape$" beampattern. In particular, the power gain is maximized at Bob's coordinate and minimized at the different Willies' coordinates simultaneously. In the case of the low correlation channel, this focal point is a single-maximum value in the range-angle distribution diagram. However, in the case of the high correlation channel, it is clear that the power gain is scattered on the beampattern, and only a partial power gain is concentrated at Bob's coordinate. Thus, this means that highly correlated legitimate and illegitimate channels can degrade the covert performance achieved by MFDA.

\subsection{MFDA with Imperfect CSI at Willies}

In this subsection, we consider the case of imperfect CSI at Willies and investigate the effects of the uncertainties in angle and range on covert rate. Larger uncertainty needs more discrete samples to evaluate covert performance which makes the optimization more complex. Hence, we only consider three different Willies for simplicity, i.e., K=3. They are located at $\left( {{r_{{w_1}}},{\theta _{{w_1}}}} \right) = \left( {1100{\rm{m}},{{10}^\circ }} \right)$, $\left( {{r_{{w_2}}},{\theta _{{w_2}}}} \right) = \left( {850{\rm{m}},{{20}^\circ }} \right)$ and $\left( {{r_{{w_3}}},{\theta _{{w_3}}}} \right) = \left( {1150{\rm{m}},{{40}^\circ }} \right)$, respectively.

\begin{figure}[htbp]
	\centering
	\begin{minipage}{0.49\linewidth}
		\centering
		\subfigure[BSUM algorithm for APV]{\includegraphics[width=1.05\textwidth]{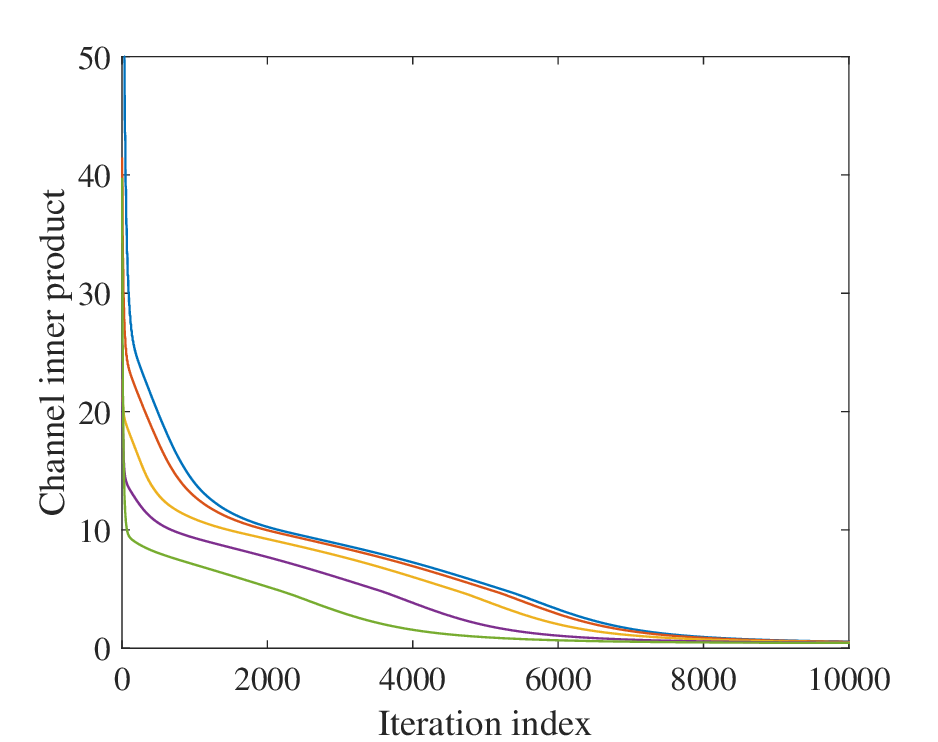}}
	\end{minipage}
	\begin{minipage}{0.49\linewidth}
		\centering
		\subfigure[BSUM algorithm for AFV]{\includegraphics[width=1.1\textwidth]{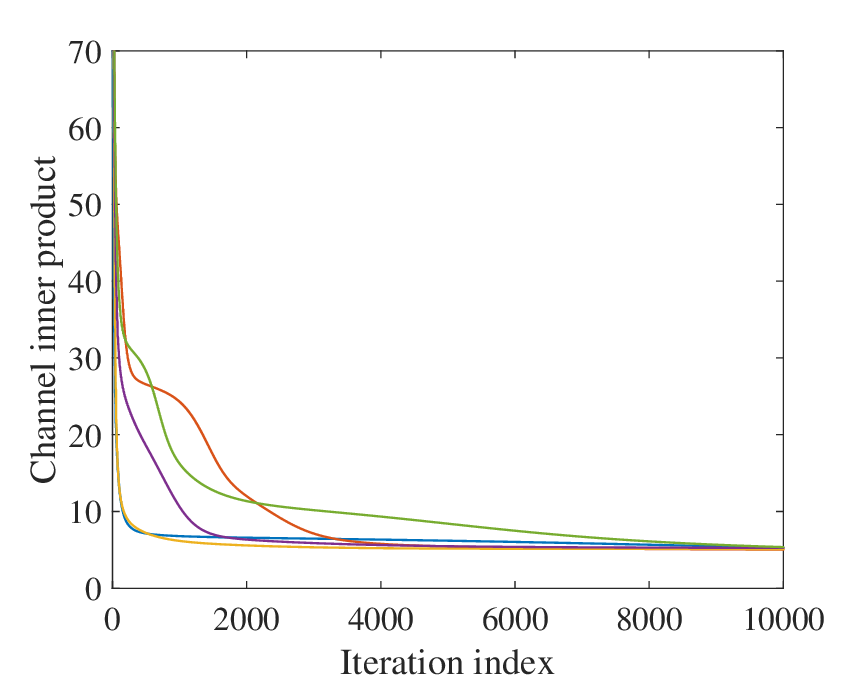}}
	\end{minipage}
\caption{Typical converging traces of the BSUM for optimizing APV and AFV with $M=10$ and $L=3$. $\Delta {r_{{w_k}}} = 10 \rm{m}$ and $\Delta {\theta _{{w_k}}} = {1^ \circ }$, k=1,...,K.}
\label{fig6}
\end{figure}

Figure 6 demonstrates the convergence of the BSUM algorithm for optimizing APV and AFV where each trace starts from random initial APV and AFV. The channel inner product is denoted as $\sum\limits_{k = 1}^K {\sum\limits_{i = 1}^L {\sum\limits_{j = 1}^L {{{\left| {\left\langle {{{\bf{h}}_{ab}},{{\bf{h}}_{a{w_k}}}\left( {{r_{k,i}},{\theta _{k,j}}} \right)} \right\rangle } \right|}^2}} } } $ in this case. Compared with Fig. 2, BSUM takes more iterations, about 8000, to converge due to the imperfect CSI at Willies. Besides, the channel inner products do not converge to zero by optimizing APV and AFV. This implies that larger uncertainties in angle and range will increase the burden of optimizing APV and AFV via BSUM.

\begin{figure}[ht]
\centering
{\includegraphics[width=0.53\textwidth]{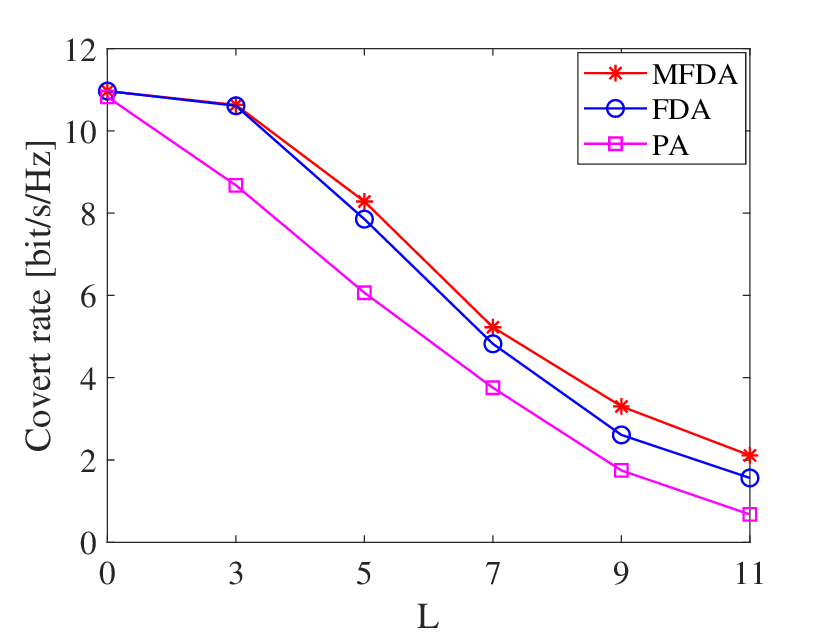}}
\caption{Covert rate $R_b$ versus number of antennas $M$.}
\label{Fig7}
\end{figure}

Figure 7 depicts the covert rate $R_b$ versus the number of antennas $M$. The uncertain regions in angle and range are denoted as ${R_{{w_k}}} \buildrel \Delta \over = \left[ {{r_{{w_k}}} - \frac{{L - 1}}{2}\Delta r_{{w_k}}^{'},{r_{{w_k}}} + \frac{{L - 1}}{2}\Delta r_{{w_k}}^{'}} \right]$ and ${\theta _{{w_k}}} \buildrel \Delta \over = \left[ {{\theta _{{w_k}}} - \frac{{L - 1}}{2}\Delta \theta _{{w_k}}^{'},{\theta _{{w_k}}} + \frac{{L - 1}}{2}\Delta \theta _{{w_k}}^{'}} \right]$, respectively. $\Delta r_{{w_k}}^{'} = 10 {\rm{m}}$ and $\Delta \theta _{{w_k}}^{'} = {1^ \circ }$ reflect the accuracy of discrete samples. To ensure that the signal power cannot be detected by any samples within the uncertain regions, MFDA needs to decrease the correlations of the legitimate channel and all possible illegitimate channels by optimizing AFV and APV. However, the uncertain regions of different Willies increase with $L$, making it increasingly difficult to decrease the correlations, which leads to a degradation in covert performance. As expected, MFDA can effectively decrease the performance loss caused by imperfect Willies CSI by utilizing the frequency dimension and the position dimension compared to FDA and PA.

\begin{figure}[htbp]
	\centering
	\begin{minipage}{0.45\linewidth}
		\centering
		\subfigure[Same uncertainty, $\Delta {r_{{w_k}}}=20 \rm{m} $ and $\Delta {\theta _{{w_k}}} = {2^ \circ }$, k=1,...,3.]{\includegraphics[width=1.0\textwidth]{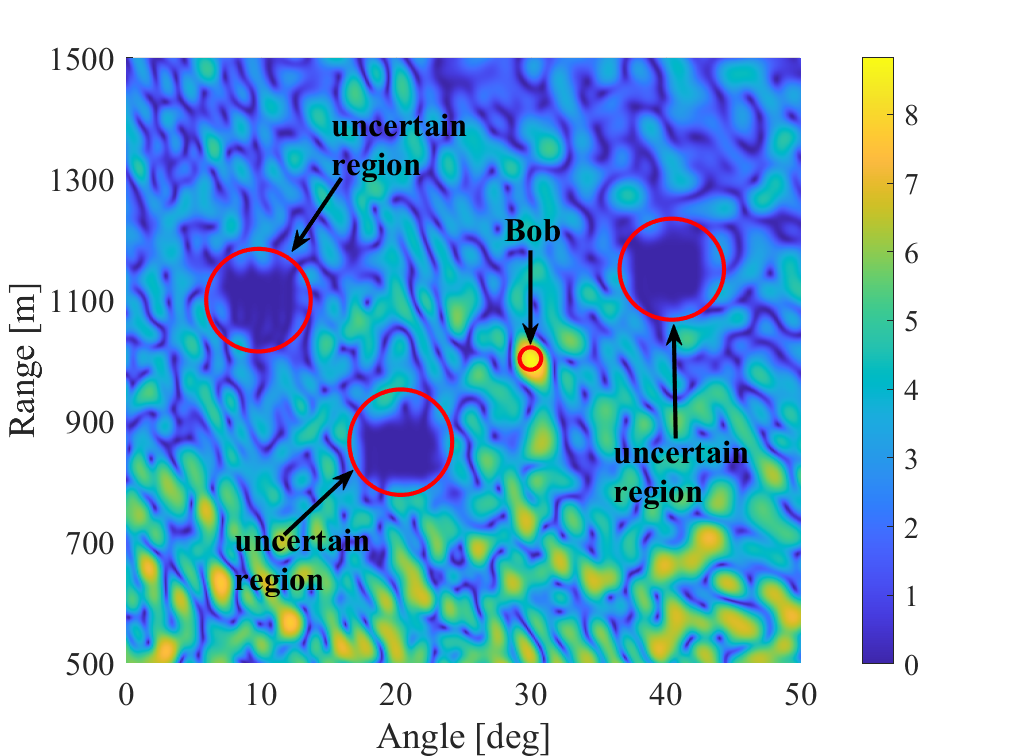}}
	\end{minipage}
	\begin{minipage}{0.45\linewidth}
		\centering
		\subfigure[Different uncertainty, $\Delta {r_{{w_k}}} = k \times 10{\rm{m}}$ and $\Delta {\theta _{{{\rm{w}}_k}}}{ = k} \times {{\rm{1}}^\circ }$, k=1,...,3.]{\includegraphics[width=1.0\textwidth]{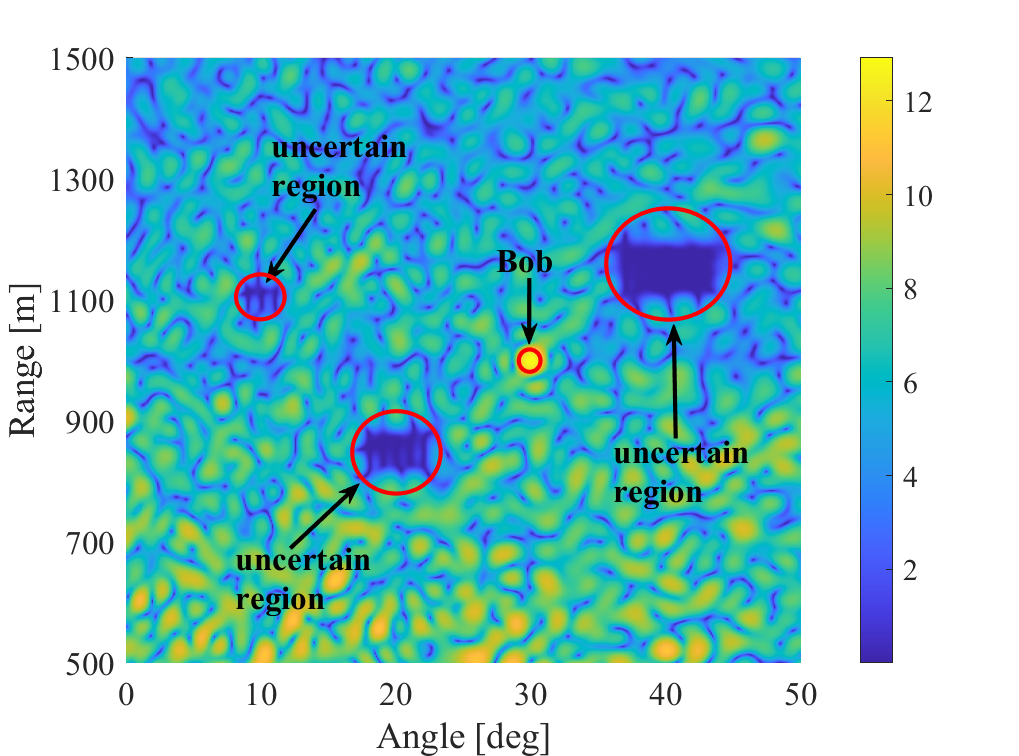}}
	\end{minipage}
\caption{Typical beampatterns for MFDA with $M=20$.}
\label{fig8}
\end{figure}

Figure 8 illustrates the beampatterns achieved by MFDA. We observe that MFDA can maximize the received rate at Bob, while ensuring that Willies in the uncertain area cannot detect the signal power. Moreover, MFDA can achieve the regional covert communication by jointly optimizing ABV, AFV and APV based on the sizes of different uncertain regions.

\section{Conclusion}
In this paper, we proposed a novel antenna technology MFDA to enhance covert performance in covert communication with multiple wardens. Specifically, we maximized the covert rate by jointly optimizing the ABV, APV and AFV at the transmitter, subject to the given constraints. Then, a two-stage AO algorithm with lower complexity was developed to solve the optimization problem. Moreover, we considered the cases of perfect covertness and imperfect CSI at wardens. Simulation results demonstrated that MFDA introduces a new dimension which can further improve covert performance based on conventional FDA. Moreover, MFDA has a strong channel identification ability to decouple the highly correlated channel, and can effectively reduce the performance loss caused by CSI estimation error.

\begin{appendices}

\section{Proof of Lemma 1 }
\subsection{Case 1}

In this case, we have $y_{k,m,n}^{'}{\left( {x_m^{s - 1};x_n^{s - 1}} \right)} > 0$, ${y_{k,m,n}}\left( {{B_{k,m,n}};x_n^{s - 1}} \right) =  - 1$ and ${B_{k,m,n}} < x_m^{s - 1}$. The relation $-1 < {y_{k,m,n}}{\left( {x_m^{s - 1};x_n^{s - 1}} \right)} < 1$ holds if and only if there exists an integer ${\kappa _1} \ge 0$ such that we have \eqref{Lemma1_case1} where $\left( a \right)$ is due to ${B_{k,m,n}} < x_m^{s - 1}$ and $\left( b \right)$ is because that ${\kappa _1}$ is an integer solution. Owing to ${y_{k,m,n}}\left( {{B_{k,m,n}};x_n^{s - 1}} \right) =  - 1$, we have

\begin{figure*}[ht]
\begin{align}\label{Lemma1_case1}\
\begin{array}{*{20}{l}}
{ - 1 < \cos \left[ {2\pi \left( {\begin{array}{*{20}{l}}
{\left( {{f_m}{x_m} - {f_n}{x_n}} \right)\frac{{\sin {\theta _{{w_k}}} - \sin {\theta _b}}}{c} + \left( {{f_m} - {f_n}} \right)\frac{{{r_b} - {r_{{w_k}}}}}{c}}
\end{array}} \right)} \right] < 1}\\
{\begin{array}{*{20}{l}}
{\mathop  \Rightarrow \limits^{\left( a \right)} 2\pi {\kappa _1} + \pi  < 2\pi \left( {\left( {{f_m}x_m^{s - 1} - {f_n}x_n^{s - 1}} \right)\frac{{\sin {\theta _{{w_k}}} - \sin {\theta _b}}}{c} + \left( {{f_m} - {f_n}} \right)\frac{{{r_b} - {r_{{w_k}}}}}{c}} \right) < 2\pi {\kappa _1} + 2\pi }\\
{ \Rightarrow \left( \begin{array}{l}
\left( {{f_m}x_m^{s - 1} - {f_n}x_n^{s - 1}} \right)\frac{{\sin {\theta _{{w_k}}} - \sin {\theta _b}}}{c}\\
 + \left( {{f_m} - {f_n}} \right)\frac{{{r_b} - {r_{{w_k}}}}}{c}
\end{array} \right) - 1 < {\kappa _1} < \left( \begin{array}{l}
\left( {{f_m}x_m^{s - 1} - {f_n}x_n^{s - 1}} \right)\frac{{\sin {\theta _{{w_k}}} - \sin {\theta _b}}}{c}\\
 + \left( {{f_m} - {f_n}} \right)\frac{{{r_b} - {r_{{w_k}}}}}{c}
\end{array} \right) - \frac{1}{2}}\\
{\mathop  \Rightarrow \limits^{\left( b \right)} {\kappa _1} = \left\lceil {\left( {{f_m}x_m^{s - 1} - {f_n}x_n^{s - 1}} \right)\frac{{\sin {\theta _{{w_k}}} - \sin {\theta _b}}}{c} + \left( {{f_m} - {f_n}} \right)\frac{{{r_b} - {r_{{w_k}}}}}{c}} \right\rceil  - 1,}
\end{array}}
\end{array}
\end{align}
\end{figure*}

\begin{align}\label{Lemma_xi1}\
2\pi \left( \begin{array}{l}
\left( {{f_m}{B_{k,m,n}} - {f_n}x_n^{s - 1}} \right)\frac{{\sin {\theta _{{w_k}}} - \sin {\theta _b}}}{c}\\
 + \left( {{f_m} - {f_n}} \right)\frac{{{r_b} - {r_{{w_k}}}}}{c}
\end{array} \right) = \pi  + 2{\kappa _1}\pi.
\end{align}
Thus, ${B_{k,m,n}}$ is given by
\begin{align}\label{Lemma_xi_case1}\
{B_{k,m,n}} &= \frac{c}{{{f_m}\left( {\sin {\theta _{{w_k}}} - \sin {\theta _b}} \right)}}\nonumber\\
 &\times \left( {\frac{{1 + 2{\kappa _1}}}{2} - \left( {{f_m} - {f_n}} \right)\frac{{{r_b} - {r_{{w_k}}}}}{c}} \right) + \frac{{{f_n}x_n^{s - 1}}}{{{f_m}}}.
\end{align}
According to \eqref{newobject_constraints}, we can obtain ${k_{m,n}}$ and ${\delta _{m,n}}$ through simple mathematical manipulations which are given by
\begin{align}\label{Lemma_case1_k}\
{A_{k,m,n}} &= \frac{{ - \pi {f_m}\frac{{\sin {\theta _{{w_k}}} - \sin {\theta _b}}}{c}}}{{x_m^{s - 1} - {B_{k,m,n}}}}\nonumber\\
 &\times \sin \left[ {2\pi \left( \begin{array}{l}
\left( {{f_m}x_m^{s - 1} - {f_n}x_n^{s - 1}} \right)\frac{{\sin {\theta _{{w_k}}} - \sin {\theta _b}}}{c}\\
 + \left( {{f_m} - {f_n}} \right)\frac{{{r_b} - {r_{{w_k}}}}}{c}
\end{array} \right)} \right]
\end{align}
and
\begin{align}\label{Lemma_case1_delta}\
{C_{k,m,n}} &= \cos \left[ {2\pi \left( \begin{array}{l}
\left( {{f_m}x_m^{s - 1} - {f_n}x_n^{s - 1}} \right)\frac{{\sin {\theta _{{w_k}}} - \sin {\theta _b}}}{c}\\
 + \left( {{f_m} - {f_n}} \right)\frac{{{r_b} - {r_{{w_k}}}}}{c}
\end{array} \right)} \right]\nonumber\\
 &- {A_{k,m,n}}{\left( {x_m^{s - 1} - {B_{k,m,n}}} \right)^2}.
\end{align}

\subsection{Case 2}
In this case, we have $y_{k,m,n}^{'}{\left( {x_m^{s - 1};x_n^{s - 1}} \right)} > 0$, ${y_{k,m,n}}\left( {{B_{k,m,n}};x_n^{s - 1}} \right) =  - 1$ and $ x_m^{s - 1} < {B_{k,m,n}}$. The relation $-1 < {y_{k,m,n}}{\left( {x_m^{s - 1};x_n^{s - 1}} \right)} < 1$ holds if and only if there exists an integer ${\kappa _2} \ge 0$ such that we have \eqref{case2}. According to \eqref{Lemma1_case1}, we note that ${\kappa _1} = {\kappa _2}$. Hence, we can obtain ${A_{k,m,n}}$, ${B_{k,m,n}}$ and ${C_{k,m,n}}$ as in \eqref{Lemma_case1_k}, \eqref{Lemma_xi_case1} and \eqref{Lemma_case1_delta}, respectively.

\begin{figure*}[ht]
\begin{align}\label{case2}\
\begin{array}{*{20}{l}}
{ - 1 < \cos \left[ {2\pi \left( {\begin{array}{*{20}{l}}
{\left( {{f_m}{x_m} - {f_n}{x_n}} \right)\frac{{\sin {\theta _{{w_k}}} - \sin {\theta _b}}}{c} + \left( {{f_m} - {f_n}} \right)\frac{{{r_b} - {r_{{w_k}}}}}{c}}
\end{array}} \right)} \right] < 1}\\
{\begin{array}{*{20}{l}}
{\mathop  \Rightarrow \limits^{\left( a \right)} 2\pi {\kappa _2} < 2\pi \left( {\left( {{f_m}x_m^{s - 1} - {f_n}x_n^{s - 1}} \right)\frac{{\sin {\theta _{{w_k}}} - \sin {\theta _b}}}{c} + \left( {{f_m} - {f_n}} \right)\frac{{{r_b} - {r_{{w_k}}}}}{c}} \right) < 2\pi {\kappa _2} + \pi }\\
{ \Rightarrow \left( \begin{array}{l}
\left( {{f_m}x_m^{s - 1} - {f_n}x_n^{s - 1}} \right)\frac{{\sin {\theta _{{w_k}}} - \sin {\theta _b}}}{c}\\
 + \left( {{f_m} - {f_n}} \right)\frac{{{r_b} - {r_{{w_k}}}}}{c}
\end{array} \right) - \frac{1}{2} < {\kappa _2} < \left( \begin{array}{l}
\left( {{f_m}x_m^{s - 1} - {f_n}x_n^{s - 1}} \right)\frac{{\sin {\theta _{{w_k}}} - \sin {\theta _b}}}{c}\\
 + \left( {{f_m} - {f_n}} \right)\frac{{{r_b} - {r_{{w_k}}}}}{c}
\end{array} \right)}\\
{\mathop  \Rightarrow {\kappa _2} = \left\lfloor {\left( {{f_m}x_m^{s - 1} - {f_n}x_n^{s - 1}} \right)\frac{{\sin {\theta _{{w_k}}} - \sin {\theta _b}}}{c} + \left( {{f_m} - {f_n}} \right)\frac{{{r_b} - {r_{{w_k}}}}}{c}} \right\rfloor ,}
\end{array}}
\end{array}
\end{align}
\end{figure*}

\subsection{Case 3}

In this case, we have $y_{k,m,n}^{'}{\left( {x_m^{s - 1};x_n^{s - 1}} \right)} =0$ and ${y_{k,m,n}}{\left( {x_m^{s - 1};x_n^{s - 1}} \right)} = -1$ which means that ${B_{k,m,n}} = x_m^{s - 1}$ and ${C_{k,m,n}} =  - 1$. To find ${A_{k,m,n}}$, we first introduce a small parameter $\kappa _3$ whose value approaches zero, i.e., ${\kappa _3} \to 0$. Then, at the point ${x_m^{s - 1} + {\kappa _3}}$, we have $y_{k,m,n}^{'}\left( {x_m^{s - 1} + {\kappa _3};x_n^{s - 1}} \right) = u_{k,m,n}^{'}\left( {x_m^{s - 1} + {\kappa _3};x_n^{s - 1}} \right)$. We can obtain
\begin{align}\label{minimumpoint2}\
\begin{array}{*{20}{l}}
{{A_{k,m,n}} = \frac{{ - \pi {f_m}\frac{{\sin {\theta _{{w_k}}} - \sin {\theta _b}}}{c}}}{{{\kappa _3}}}}\\
{ \times \sin \left[ {2\pi \left( {\begin{array}{*{20}{l}}
{\left( {{f_m}\left( {x_m^{s - 1} + {\kappa _3}} \right) - {f_n}x_n^{s - 1}} \right)}\\
{ \times \frac{{\sin {\theta _{{w_k}}} - \sin {\theta _b}}}{c}}\\
{ + \left( {{f_m} - {f_n}} \right)\frac{{{r_b} - {r_{{w_k}}}}}{c}}
\end{array}} \right)} \right]}\\
{\mathop  \Rightarrow \limits^{\left( c \right)} \frac{{\pi {f_m}\frac{{\sin {\theta _{{w_k}}} - \sin {\theta _b}}}{c}\sin \left( {2\pi \frac{{\sin {\theta _{{w_k}}} - \sin {\theta _b}}}{c}{f_m}{\kappa _3}} \right)}}{{{\kappa _3}}}.}
\end{array}
\end{align}
where $\left( c \right)$ is due to the sum difference of product method \cite{IS}. It is worth noting that $\mathop {\lim }\limits_{{\kappa _3} \to 0} \frac{{\pi {f_m}\frac{{\sin {\theta _{{w_k}}} - \sin {\theta _b}}}{c}\sin \left( {2\pi \frac{{\sin {\theta _{{w_k}}} - \sin {\theta _b}}}{c}{f_m}{\kappa _3}} \right)}}{{{\kappa _3}}} = 2{\left( {\pi {f_m}\frac{{\sin {\theta _{{w_k}}} - \sin {\theta _b}}}{c}} \right)^2}$, thus we have ${A_{k,m,n}} = 2{\left( {\pi {f_m}\frac{{\sin {\theta _{{w_k}}} - \sin {\theta _b}}}{c}} \right)^2}$.

\subsection{Case 4}
In this case, we have $y_{k,m,n}^{'}{\left( {x_m^{s - 1};x_n^{s - 1}} \right)} =0$ and ${y_{k,m,n}}\left( {x_m^{s - 1};x_n^{s - 1}} \right) = 1$ which means that ${B_{k,m,n}} = x_m^{s - 1}$ and ${C_{k,m,n}} =  1$. Similar to case 3, we have $y_{k,m,n}^{'}\left( {x_m^{s - 1} + {\kappa _3};x_n^{s - 1}} \right) = u_{k,m,n}^{'}\left( {x_m^{s - 1} + {\kappa _3};x_n^{s - 1}} \right)$, such that we have ${A_{k,m,n}} =  - 2{\left( {\pi {f_m}\frac{{\sin {\theta _{{w_k}}} - \sin {\theta _b}}}{c}} \right)^2}$.

\section{}

According to \eqref{covert_constraintopspecia2}, ${\bf{w}}\left( t \right)$ needs to lie in the null space of ${{\bf{H}}_{{\bf{aw}}}}$. Thus, it takes the form ${\bf{w}} = \left( {{I_M} - {\bf{H}}_{{\bf{aw}}}^\dag {{\left( {{{\bf{H}}_{{\bf{aw}}}}{\bf{H}}_{{\bf{aw}}}^\dag } \right)}^{ - 1}}{{\bf{H}}_{{\bf{aw}}}}} \right){{\bf{w}}_0}$ where ${{\bf{w}}_0} \in {{\mathbb{C}}^M}$. Next, we note that ${\left| {{{\bf{h}}_{ab}}{\bf{w}}} \right|^2} = {\left| {{{\bf{h}}_{ab}}\left( {{I_M} - {\bf{H}}_{{\bf{aw}}}^\dag {{\left( {{{\bf{H}}_{{\bf{aw}}}}{\bf{H}}_{{\bf{aw}}}^\dag } \right)}^{ - 1}}{{\bf{H}}_{{\bf{aw}}}}} \right){{\bf{w}}_0}} \right|^2} \le {\left\| {{{\bf{h}}_{ab}}\left( {{I_M} - {\bf{H}}_{{\bf{aw}}}^\dag {{\left( {{{\bf{H}}_{{\bf{aw}}}}{\bf{H}}_{{\bf{aw}}}^\dag } \right)}^{ - 1}}{{\bf{H}}_{{\bf{aw}}}}} \right)} \right\|^2}\left\| {{{\bf{w}}_0}} \right\|$, and the equality holds when ${{{\bf{h}}_{ab}}\left( {{I_M} - {\bf{H}}_{{\bf{aw}}}^\dag {{\left( {{{\bf{H}}_{{\bf{aw}}}}{\bf{H}}_{{\bf{aw}}}^\dag } \right)}^{ - 1}}{{\bf{H}}_{{\bf{aw}}}}} \right)}$ and ${{{\bf{w}}_0}}$ are collinear. Note that ${\left\| {\bf{w}} \right\|^2} \le {P_{max}}$, we can obtain that ${{\bf{w}}_0} = \frac{{\sqrt {{P_{\max }}} {\bf{h}}_{ab}^\dag }}{{\left\| {\left( {{I_M} - {\bf{H}}_{{\bf{aw}}}^\dag {{\left( {{{\bf{H}}_{{\bf{aw}}}}{\bf{H}}_{{\bf{aw}}}^\dag } \right)}^{ - 1}}{{\bf{H}}_{{\bf{aw}}}}} \right){\bf{h}}_{ab}^\dag } \right\|}}$. The optimal ABV can be written as ${{\bf{w}}_{opt}} = \sqrt {{P_{\max }}} \frac{{\left( {{I_M} - {\bf{H}}_{{\bf{aw}}}^\dag {{\left( {{{\bf{H}}_{{\bf{aw}}}}{\bf{H}}_{{\bf{aw}}}^\dag } \right)}^{ - 1}}{{\bf{H}}_{{\bf{aw}}}}} \right){\bf{h}}_{ab}^\dag }}{{\left\| {\left( {{I_M} - {\bf{H}}_{{\bf{aw}}}^\dag {{\left( {{{\bf{H}}_{{\bf{aw}}}}{\bf{H}}_{{\bf{aw}}}^\dag } \right)}^{ - 1}}{{\bf{H}}_{{\bf{aw}}}}} \right){\bf{h}}_{ab}^\dag } \right\|}}$.

\end{appendices}

\bibliography{FDA_MA}

\end{document}